\documentclass[fleqn,useAMS]{Papers}
\usepackage{newtxtext,newtxmath}
\usepackage[T1]{fontenc}
\usepackage{subfloat}
\usepackage{makecell}

\usepackage[square,sort,numbers]{natbib}

\usepackage{graphicx}	
\usepackage{amsmath}	
\usepackage{lipsum}
\usepackage{caption}
\usepackage{comment}
\usepackage{textgreek} 
\usepackage{xcolor}  
\newcommand{\rev}[1]{{\color{black} #1}} 

\title[Critical acceleration of dark matter]{On a critical acceleration scale of dark matter in \texorpdfstring{$\Lambda$}{}CDM and dynamical dark energy}

 \author[Z. Xu]{Zhijie (Jay) Xu,$^{1}$\thanks{E-mail: \href{mailto:zhijie.xu@pnnl.gov}{zhijie.xu@pnnl.gov}; \href{mailto:zhijiexu@hotmail.com}{zhijiexu@hotmail.com}}
\\
$^{1}$Physical and Computational Sciences Directorate, Pacific Northwest National Laboratory; Richland, WA 99354, USA\\
}

\date{Accepted XXX. Received YYY; in original form ZZZ}

\pubyear{2023}

\begin{document}
\label{firstpage}
\pagerange{\pageref{firstpage}--\pageref{lastpage}}
\maketitle

\begin{abstract}
\noindent 
Universal acceleration $a_0$ emerges in various empirical laws, yet its fundamental nature remains unclear. Using Illustris and Virgo N-body simulations, we focus on the velocity and acceleration fluctuations in collisionless dark matter involving long-range gravity. In contrast, in the kinetic theory of gases, molecules undergo random elastic collisions involving short-range interactions, where only velocity fluctuations are relevant. 
Hierarchical structure formation proceeds through the merging of smaller haloes to form larger haloes, which facilitates a continuous energy cascade from small to large haloes at a constant rate $\varepsilon_u\approx -10^{-7}$m$^2$/s$^3$. Velocity fluctuations involve a critical velocity $u_c\propto (1+z)^{-3/4}$. Acceleration fluctuations involve a critical acceleration $a_c\propto (1+z)^{3/4}$. Two critical quantities are related by the rate of energy cascade $\varepsilon_{u}\approx -{a_c u_c/[2(3\pi)^2]}$, where factor $3\pi$ is from the angle of incidence during merging. With critical velocity $u_c$ on the order of 300 km/s at $z=0$, the critical acceleration is determined to be $a_{c0}\equiv a_c(z=0) \approx 10^{-10}$m/s$^2$, suggesting $a_c$ might explain the universal acceleration $a_0\approx 10^{-10}$m/s$^2$ in the empirical Tully-Fisher relation or modified Newtonian dynamics (MOND). The redshift evolution $a_c \propto (1+z)^{3/4}$ is in good agreement with Magneticum and EAGLE simulations and in reasonable agreement with limited observations. This suggests a larger $a_0$ at a higher redshift such that galaxies of fixed mass rotate faster at a higher redshift. Note that dark energy (DE) density $\rho_{DE0}\approx {a_{c0}^{2}/G}=10^{-10}$J/m$^3$, we postulate an entropic origin of the dark energy from acceleration fluctuations of dark matter, in analogy to the gas pressure from velocity fluctuations. This leads to a dynamical dark energy coupled to the structure evolution involving a relatively constant DE density followed by a slow weakening phase, suggesting possible deviations from the standard $\Lambda$CDM paradigm. 
\end{abstract}

\begin{keywords}
\vspace*{-10pt}
Dark matter; Tully-Fisher; Galaxy rotation; MOND; Dark energy;
\end{keywords}

\begingroup
\let\clearpage\relax
\tableofcontents
\endgroup
\vspace*{-20pt}

\section{Introduction}
\label{sec:1}
The dark matter problem originates from the mass discrepancy between the amount of dynamical mass required by the motions of astronomical objects and the directly observed luminous mass. The flat rotation curves of spiral galaxies directly point to that mass discrepancy problem: the total mass predicted by Newtonian gravity is much greater than the observed luminous mass \citep{Rubin:1970-Rotation-of-Andromeda-Nebula-f,Rubin:1980-Rotational-Properties-of-21-Sc}. The standard cosmological model ($\Lambda$CDM) interprets this mass discrepancy in terms of cold dark matter (CDM) \citep{Peebles:1984-Tests-of-cosmological-models,Spergel:2003-First-Year-Wilkinson-Microwave-Anisotropy,Komatsu:Seven-year-Wilkinson-Microwave-Anisotropy-Probe} that is believed to be cold (non-relativistic), collisionless, dissipationless, non-baryonic, and interacts with baryonic matter only through gravity. The $\Lambda$CDM model has been highly successful in the formation and evolution of large-scale structures, the cosmic microwave background, the distribution of galaxies, and the matter content of the universe \citep{Peebles:1984-Tests-of-cosmological-models,Spergel:2003-First-Year-Wilkinson-Microwave-Anisotropy,Komatsu:Seven-year-Wilkinson-Microwave-Anisotropy-Probe,Frenk:2012-Dark-matter-and-cosmic-structure}. However, the theory also encounters several difficulties on small scales \citep{Bullock:2017-Small-Scale-Challenges-to-the,DelPopolo:2017-Small-scale-problems-of-the,Perivolaropoulos:2022-Challenges-for} in describing small-scale structures, including the "cusp/core" problem \citep{Flores:1994-Observational-and-Theoretical-Constraints,Moore:1994-Evidence-against-dissipation-less-dark-matter,deBlok:2009-The-Core-Cusp-Problem}, the "missing satellite" problem \citep{Klypin:1999-Where-Are-the-Missing-Galactic,Moore:1999-Dark-Matter-Substructure,Boylan_Kolchin:2011-Too-big-to-fail,Boylan-Kolchin:2012-The-Milky-Ways-bright-satellites}, and the problem of re-obtaining the baryonic Tully-Fisher relation \citep{DelPopolo:2017-Small-scale-problems-of-the}. 

The original Tully-Fisher relation was proposed to empirically correlate the rotation velocity $v_f$ of a spiral galaxy with its intrinsic luminosity or stellar mass \citep{Tully:1977-New-Method-of-Determining-Dist}. A similar scaling was also observed for the dispersion of the random velocity of stars \citep{Faber:1976-Velocity-Dispersions-and-Mass-}. The Tully-Fisher relation has been generalized to using the total baryonic mass $M_b$ instead of just the stellar mass. The resulting baryonic Tully-Fisher relation (BTFR) is a well-established empirical relation between the total baryonic mass $M_b$ (stellar plus gas) and the flat rotation velocity $v_f$ that extends for six decades in $M_b$ \citep{McGaugh:2000-The-baryonic-Tully-Fisher-rela,Bell:2001-Stellar-Mass-to-Light-Ratios,Verheijen:2001-The-Ursa-Major-Cluster-of-Galaxies}. In BTFR, the scaling $v_{f} \propto M_b^{{1/4}}$ between the rotation velocity $v_f$ and the baryonic mass $M_b$ \citep{Wheeler:2019-The-Radial-Acceleration-Relation} can be written as 
\begin{equation} 
\label{eq:6-1} 
v_{f}^{4}=G M_b a_{0}, 
\end{equation}
where $G$ is the gravitational constant and $a_{0} \approx 10^{-10}$m/s$^2$ is an empirical constant of acceleration. 

One problem in understanding the BTFR lies in the discrepancy between the scatter and the slope inferred from the model and those from observations \citep{Dutton:2012-The-Baryonic-Tully-Fisher-relation}. 
Furthermore, the BTFR obtained from $\Lambda$CDM simulations also shows a larger scatter than that of observations \citep{Dutton:2012-The-Baryonic-Tully-Fisher-relation}. Some recent simulations that account for stellar feedback processes have shown improved agreement. However, it remains challenging to understand the scaling and scatter of the Tully-Fisher relation \citep{DelPopolo:2017-Small-scale-problems-of-the,Famaey:2013-Challenges-fo-CDM-and-MOND}, as well as the origin of empirical acceleration $a_0$.

Beyond the $\Lambda$CDM cosmology, an alternative interpretation resorts to the modification of gravity that might eliminate the need for dark matter for the mass discrepancy problem. The modified Newtonian dynamics (MOND) is an empirical model proposed to reproduce these astronomical observations without invoking the dark matter hypothesis \citep{Milgrom:1983-A-Modification-of-the-Newtonia}. The basic idea of MOND is to introduce a universal acceleration scale $a_{0} \approx 1.2\times 10^{-10} {m/s^{2} } $. Standard Newtonian mechanics $F=ma$ is recovered when acceleration $a\gg a_{0} $. For the so-called "deep-MOND" regime with $a\ll a_{0}$, Newtonian mechanics needs to be modified to $F=m{a^{2}/a_0} $, that is, the external force $F\propto a^2$. Section \ref{sec:8} provides more details on this ad hoc theory. MOND might explain the shape of the rotation curves \citep{McGaugh:1998-Testing-the-dark-matter-hypoth}, the baryonic Tully-Fisher relation \citep{McGaugh:2000-The-baryonic-Tully-Fisher-rela,Lelli:2019-The-baryonic-Tully-Fisher-rela}, and the tight correlation between the radial acceleration of the rotation curves and that of the observed baryons \citep{McGaugh:2016-Radial-Acceleration-Relation}. However, recent studies also suggest that $\Lambda$CDM is also consistent with the radial acceleration relation \citep{Keller:2017-CDM-is-Consistent-with-SPARC}. This motivates the search for a theory for the observed scaling ($v_{f} \propto M_b^{{1/4}}$) and the origin of the empirical acceleration $a_0$ within the $\Lambda$CDM paradigm.

\begin{figure}
\includegraphics*[width=\columnwidth]{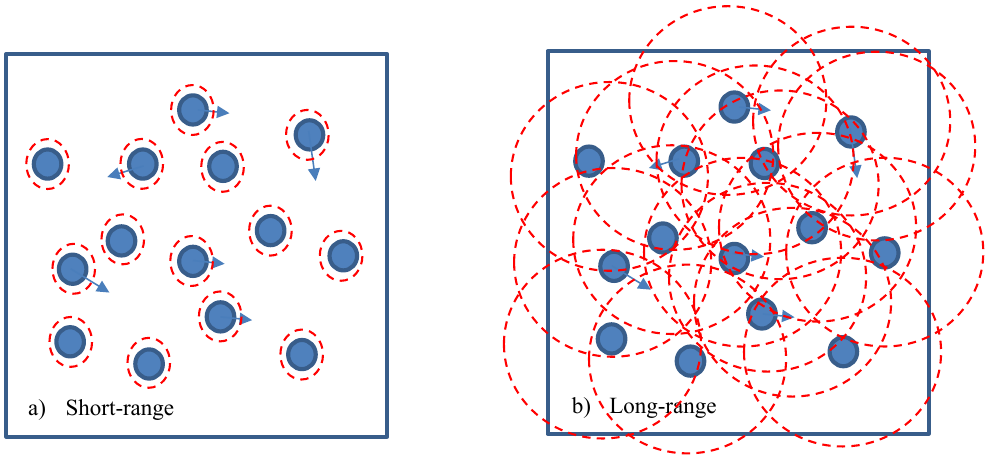}
\caption{The Schematic plot of systems that involve short-range and long-range interactions (red dashed line for interaction range): a) In the kinetic theory of gases, molecules undergo random elastic collisions that involve only short-range interactions. Molecules interact only with other molecules within a very short distance. For a dilute system, most of the molecules are out of the interaction range of other molecules. Therefore, only velocity fluctuations are relevant, which usually follows a Boltzmann distribution (Eq. \eqref{ZEqnNum8641322}); b) For self-gravitating dark matter, particles interact with all other particles through long-range gravity such that a small change in any particle position leads to perturbations in accelerations of all other particles. Therefore, both velocity fluctuations and acceleration fluctuations are relevant for self-gravitating dark matter and are discussed in detail in this paper (Fig. 
\ref{fig:4} and Eq. \eqref{ZEqnNum864132}).}
\label{fig:S4}
\end{figure}

Small-scale challenges to $\Lambda$CDM strongly indicate missing pieces in our understanding of dark matter physics. In this paper, we discuss the acceleration fluctuations in dark matter and identify a fluctuation-induced critical acceleration $a_c$ that might explain the empirical acceleration $a_0$ in BTFR. The existence of velocity fluctuations in dark matter is well known, where the probability distributions of dark matter velocity have been extensively studied \citep{Gorski:1989-Cosmological-Velocity-Correlat, Hahn:2015-The-properties-of-cosmic-veloc, Xu:2023-Maximum-entropy-distributions, Xu:2023-On-the-statistical-theory-of-self-gravitating, Xu:2024-On-the-statistical-theory-of-self-gravitating-scale_redshift}. This paper focuses on the acceleration fluctuations in dark matter and their effects on the dynamics of galaxies and dark matter haloes. Potential connections between acceleration fluctuations and dark energy were also proposed in analogy to the connections between pressure and velocity fluctuations.

As shown in Fig. \ref{fig:S4} b), due to the long-range nature of gravity, every dark matter particle interacts with all other particles such that a small change in any particle position leads to perturbations in the acceleration of all other particles. This leads to fluctuations in particle acceleration and a redshift-dependent probability distribution of particle accelerations. On the contrary, in the kinetic theory of gases in Fig. \ref{fig:S4} a), molecules undergo random elastic collisions that involve short-range interactions. Molecules interact only with other molecules within a very short distance. Changes in any molecule's position do not lead to perturbations in the acceleration of other molecules. Therefore, only velocity fluctuations are relevant for systems involving short-range interactions. 

Compared to the velocity distributions and correlations that have been extensively studied \citep{Xu:2023-Maximum-entropy-distributions, Xu:2023-On-the-statistical-theory-of-self-gravitating, Xu:2024-On-the-statistical-theory-of-self-gravitating-scale_redshift,Xu:2024-On-the-statistical-theory-of-self-gravitating}, more work is required to understand the nature of acceleration fluctuations and distributions on different length and time scales. The existence of acceleration fluctuations naturally leads to a critical acceleration scale $a_c$, the root mean square (RMS) acceleration, for a given distribution of particle accelerations. In this paper, we will focus on the acceleration fluctuations, the critical acceleration $a_c$, its relation to empirical acceleration $a_0$ in BTFR, and potential connections to dark energy. The remainder of the paper is organized as follows. Section \ref{sec:5} presents the redshift evolution of the acceleration distributions from the N-body simulations, followed by the velocity distributions in Section \ref{sec:4}. The velocity and acceleration fluctuations are closely related to each other, connected by an energy cascade process during the hierarchical structure formation (Sections \ref{sec:6-1} and \ref{sec:6}). Section \ref{sec:8-2} presents the simulations and observations for the redshift variation of $a_0$. The connection with empirical MOND acceleration $a_0$ is presented in Section \ref{sec:8}, along with a theory for "Non-Newtonian" behavior at small acceleration ($a\ll a_0$). Section \ref{sec:7} presents relevant physical quantities based on fluctuations in dark matter, followed by Section \ref{sec:7-2} on potential connections to dark energy.

\section{Distributions of dark matter acceleration}
\label{sec:5}
In this section, we focus on the acceleration distributions of dark matter particles from N-body simulations. The basic dynamics of the self-gravitating collisionless dark mater flow (SG-CFD) is governed by collisionless Boltzmann equations (CBE) \citep{Mo:2010-Galaxy-formation-and-evolution} that can be numerically solved by particle-based N-body simulations \citep{Peebles:1980-The-Large-Scale-Structure-of-t}. As a first step, we focus on the dark matter only simulations. The simulation data were generated from the large-scale N-body simulations by the Virgo consortium or the SCDM model (the matter-dominant cold dark matter model) with $\Omega_{DM}=1$. This suite of cosmological simulations is a dark matter only with a 240 Mpc$^3$ volume. Each dark matter particle has a mass around $m_p=4.54\times 10^{11} M_{\odot}$. The gravitational softening length is around 36 kpc. More details can be found in \cite{Frenk:2000-Public-Release-of-N-body-simul}. The same set of simulation data has also been widely used to study clustering statistics \citep{Jenkins:1998-Evolution-of-structure-in-cold}, the formation of halo clusters \citep{Colberg:1999-Linking-cluster-formation-to-l}, and to test halo mass functions \citep{Sheth:2001-Ellipsoidal-collapse-and-an-im}. 

In this paper, to cross-check some results, we also use the Illustris simulations (Illustris-1-Dark for dark matter only), a more recent suite of cosmological simulations with a much higher mass resolution of $m_p=7.6\times 10^6 M_{\odot}$ \citep{NELSON:2015-The-illustris-simulation}. The simulation has cosmological parameters of dark matter density $\Omega_{DM}=0.2726$, dark energy density $\Omega_{DE}=0.7274$ at $z=0$, and Hubble constant $h=0.704$. The simulation results of the Illustris simulations will be explicitly indicated in the figures. 

For both simulations, the Friends of Friends (FOF) algorithm was used to identify all haloes in the simulation that depend only on a dimensionless parameter \textit{b}, which defines the linking length $b\left({N/V} \right)^{{-1/3} }$. Here, $N$ is the number of total particles, and $V$ is the volume of the simulation box. All haloes in the simulation were identified with a linking length parameter $b=0.2$. After identifying all haloes in the N-body system, all dark matter particles can be divided into two groups: halo particles ("hp") and out-of-halo particles ("op", particles that do not belong to any haloes). This is important because the velocity and acceleration statistics of particles in two groups are very different and should be studied separately.

By calculating the total force for every dark matter particle, the proper acceleration $\boldsymbol{\mathrm{a}}_p$ for particle \textit{i} is
\begin{equation} 
\label{eq:6} 
\boldsymbol{\mathrm{a}}_{p} =\frac{Gm_{p} }{a^{2} } \sum _{j\ne i}^{N}\frac{\boldsymbol{\mathrm{x}}_{i} -\boldsymbol{\mathrm{x}}_{j} }{\left|\boldsymbol{\mathrm{x}}_{i} -\boldsymbol{\mathrm{x}}_{j} \right|^{3} }  ,         
\end{equation} 
where $\boldsymbol{\mathrm{x}}_{i} $ and $\boldsymbol{\mathrm{x}}_{j} $ are the comoving coordinates of particles \textit{i} and \textit{j}. The summation runs over all other particles except \textit{i}. The periodic boundary is also taken care of in the calculation, with a total of 26 replicas of the simulation domain in three dimensions to approximate the long-range gravity. 

\begin{figure}
\includegraphics*[width=\columnwidth]{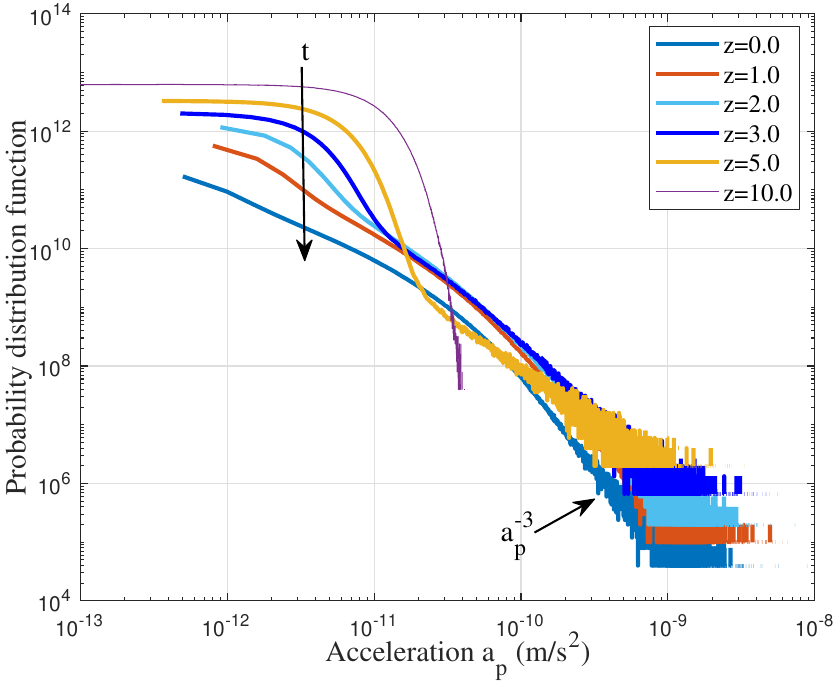}
\caption{The redshift evolution of the probability distribution of particle acceleration $a_{p}$ for all particles in an N-body system (SCDM). A long tail $\propto a_{p}^{-3} $ is gradually formed from \textit{z}=5 due to the formation of halo structures.}
\label{fig:3}
\end{figure}

Figure \ref{fig:3} plots the redshift variation of the acceleration distribution, i.e., the distribution of Cartesian components [$a_{px}, a_{py}, a_{pz} $] of the acceleration vector $\boldsymbol{\mathrm{a}}_{p} $ for all dark matter particles. Particle acceleration evolves from an initial Gaussian distribution at high redshift to a distribution with a long tail $\propto a_{p}^{-3} $ for large acceleration in the halo core region. The tail starts to form at around \textit{z}=5 due to the formation and evolution of halo structures. 

\begin{figure}
\includegraphics*[width=\columnwidth]{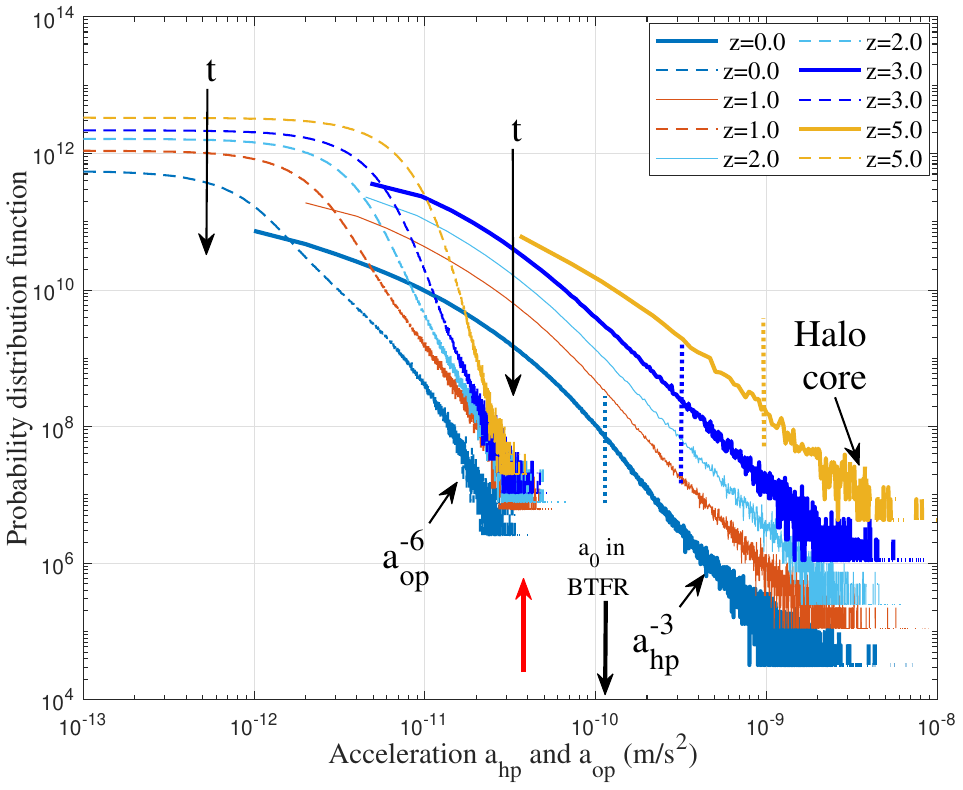}
\caption{The redshift evolution of the probability distributions of halo particle acceleration $a_{hp}$ (solid lines) and out-of-halo particle acceleration $a_{op}$ (dash lines). A long tail $\propto a_{hp}^{-3}$ at large acceleration is a typical feature due to halo particles in the core region. The distribution of $a_{op}$ for out-of-halo particles is approximately Gaussian at high redshift. For both types of particles, the acceleration decreases with time because of expanding space. The dotted lines represent the standard deviation of the acceleration distribution (RMS acceleration) at $z=$5, 3, and 0. The time variation of the RMS acceleration is presented in Fig. \ref{fig:5}. The empirical BTFR acceleration ($a_{0} \approx 10^{-10}$m/s$^2$) is marked on the graph (black arrow) that matches the RMS acceleration of the N-body simulation at $z=0$. This is not just a coincidence but suggests a fluctuation origin of the empirical acceleration $a_0$ in BTFR (Section \ref{sec:8}).}
\label{fig:4}
\end{figure}

\begin{figure}
\includegraphics*[width=\columnwidth]{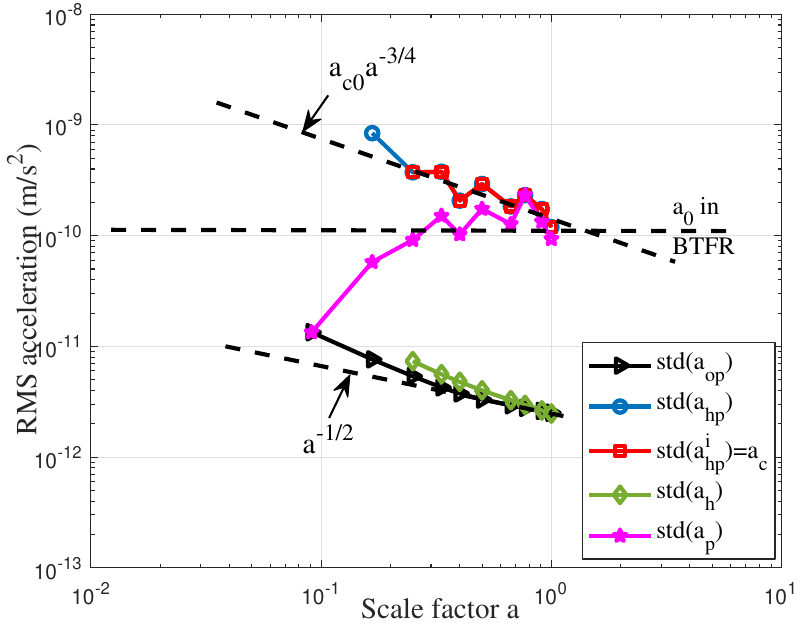}
\caption{The variation of RMS accelerations (m/s$^2$) with scale factor \textit{a} for all particles($a_p$: magenta), halo particles ($a_{hp}$: blue), out-of-halo particles ($a_{op}$: black), haloes ($a_{h}$: green), and acceleration fluctuation of all halo particles ($a_{hp}^i$: red). All accelerations decrease with time except the RMS acceleration for all particles $a_p$. The halo acceleration $a_{h}$ matches the acceleration of the out-of-halo particles $a_{op}$ and is much smaller ($\sim 10^{-12} {m/s^{2}}$) due to weaker gravity on large scales. The acceleration of the halo particles $a_{hp}$ can be decomposed into the halo acceleration $a_h$ and the fluctuation $a_{hp}^i$ (Eq. \eqref{ZEqnNum576014}). Here, std$(a_h)\propto a^{-1/2}$ evolves in the linear regime, and the critical acceleration scale $a_c$=std$(a_{hp}^i)\propto a^{-3/4}$ evolves in the nonlinear regime (Eq. \eqref{ZEqnNum138201}). At \textit{z}=0, the critical acceleration of fluctuation ($a_c$) matches the empirical acceleration $a_{0} = 10^{-10}$m/s$^2$, suggesting the connection between the critical acceleration $a_c$ and the empirical acceleration $a_0$ in the baryonic Tully-Fisher relation (BTFR in Section \ref{sec:8}). The RMS acceleration for all particles $a_p$ is approximately a weighted average of out-of-halo particle acceleration $a_{op}$ and halo particle acceleration $a_{hp}$ (Eq. \eqref{eq:25}). It does increase with time due to the increasing number of halo particles from halo structure formation and evolution.}
\label{fig:5}
\end{figure}

By dividing all particles into halo and out-of-halo particles, Fig. \ref{fig:4} plots the evolution of the acceleration distribution for halo particles ($a_{hp}$: solid lines) and for out-of-halo particles ($a_{op}$: dashed lines), respectively. The long tail $\propto a_{hp}^{-3}$ at large acceleration comes from the halo core region with a higher dark matter density. The maximum acceleration is determined by the highest density in the halo core and seems independent of the redshift. With continuous mass accretion, more particles are accreted into the halo outskirts, and the distribution gradually extends to smaller acceleration. The distribution of $a_{op}$ for out-of-halo particles is approximately Gaussian at high redshift. Acceleration decreases with time due to expanding space (also see Fig. \ref{fig:5}). The out-of-halo particles with the largest acceleration should be those particles close to the haloes with which they are going to merge (red arrow). The dotted lines represent the standard deviation of the acceleration distribution (the root mean square acceleration or RMS) at $z=$5, 3, and 0. The empirical BTFR acceleration $a_0$ is marked by the black arrow that matches the RMS acceleration at $z=0$ (blue dotted line). This is not a coincidence. We will show that the RMS acceleration, i.e. the scale of acceleration fluctuation, is related to $a_0$ in Section \ref{sec:8}.

To better describe the distribution and evolution of acceleration and velocity, we start by decomposing the particle acceleration and velocity into two parts of different natures. In N-body simulations, every halo particle ("hp"), characterized by a mass, $m_{p}$, a velocity vector, $\boldsymbol{\mathrm{v}}_{{{hp}}}$, and an acceleration vector $\boldsymbol{\mathrm{a}}_{{{hp}}}$, should belong to one and only one particular parent halo. For each identified halo, the total acceleration $\boldsymbol{\mathrm{a}}_{hp}$ and velocity $\boldsymbol{\mathrm{v}}_{hp}$ of each halo particle can be decomposed into mean values and fluctuations around these mean values \citep{Xu:2023-Maximum-entropy-distributions}. These mean values are the halo acceleration and halo velocity, respectively,
\begin{equation}
\begin{split}
\boldsymbol{\mathrm{a}}_{h} =\langle \textbf{a}_{hp} \rangle _{h}=\frac{1}{n_{p} } \sum _{k=1}^{n_{p} }\boldsymbol{\mathrm{a}}_{hp}, \boldsymbol{\mathrm{v}}_{h}  =\langle \textbf{v}_{hp} \rangle _{h}=\frac{1}{n_{p} } \sum _{k=1}^{n_{p} }\boldsymbol{\mathrm{v}}_{hp}, 
\end{split}
\label{ZEqnNum680194}
\end{equation}
\noindent where $\langle \cdot \rangle _{h}$ stands for the average of a quantity over all $n_p$ particles in a given halo. These mean values are related to the inter-halo interactions between any particle and all other particles outside the halo in which the particle resides. The inter-halo interactions are weaker, operating on larger scales and in the linear regime \citep{Xu:2022-Postulating-dark-matter-partic}. 

On the other hand, fluctuations in acceleration and velocity of halo particles are defined as 
\begin{equation}
\begin{split}
&\boldsymbol{\mathrm{a}}_{hp}^{i} =\boldsymbol{\mathrm{a}}_{hp} -\left\langle \boldsymbol{\mathrm{a}}_{hp} \right\rangle _{h} =\boldsymbol{\mathrm{a}}_{hp} -\boldsymbol{\mathrm{a}}_{h}, \\
&\boldsymbol{\mathrm{v}}_{hp}^{i} =\boldsymbol{\mathrm{v}}_{hp} -\left\langle \boldsymbol{\mathrm{v}}_{hp} \right\rangle _{h} =\boldsymbol{\mathrm{v}}_{hp} -\boldsymbol{\mathrm{v}}_{h},  
\end{split}
\label{ZEqnNum576014}
\end{equation}
where acceleration fluctuation $\boldsymbol{a}_{hp}^i$ and velocity fluctuation $\boldsymbol{v}_{hp}^i$ are related to the intra-halo interactions between that particle and all other particles in the same halo. In contrast, the intra-halo interactions are stronger, operating on smaller scales and in the nonlinear regime \citep{Xu:2022-Postulating-dark-matter-partic, Xu:2024-Cosmic-quenching-and-scaling-laws}. 

\begin{figure}
\includegraphics*[width=\columnwidth]{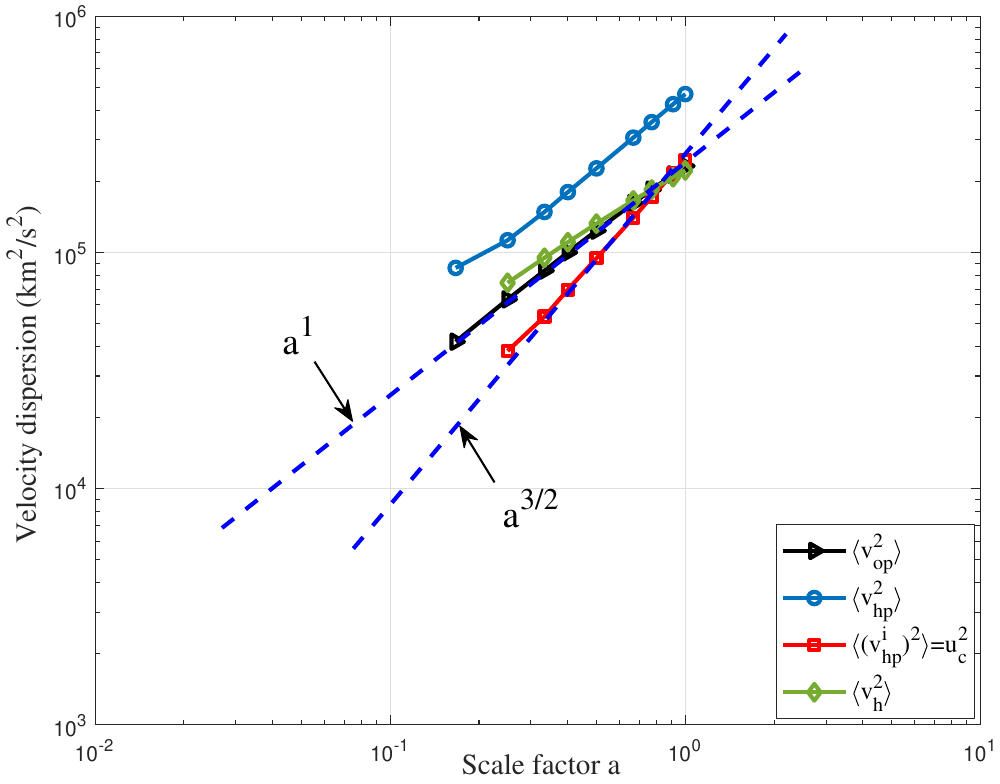}
\caption{The variation of velocity dispersions (km$^2$/s$^2$) with scale factor \textit{a} for halo particles ($v_{hp}$: blue), out-of-halo particles ($v_{op}$: black), haloes ($v_{h}$: green), and velocity fluctuation of halo particles ($v_{hp}^i$: red). All velocity dispersions increase with time. The dispersion of halo velocity $v_{h}$ matches the out-of-halo particle velocity $v_{op}$. The velocity of halo particles $v_{hp}$ can be decomposed into the halo velocity $v_h$ and the fluctuation $v_{hp}^i$ (Eq. \eqref{ZEqnNum576014}). Here, the dispersion of halo velocity $v_h$ evolves as $\propto a \propto t^{2/3}$ (in the linear regime due to inter-halo interaction on large scales). The dispersion of velocity fluctuation $v_{hp}^i$ (or the critical velocity scale $u_c^2$) evolves as $\propto a^{3/2}\propto t$ (in the nonlinear regime due to intra-halo interaction on small scales).}
\label{fig:6}
\end{figure}

Due to the different nature of inter- and intra-halo interactions, the halo velocity and acceleration ($\boldsymbol{\mathrm{v}}_{h}$ and $\boldsymbol{\mathrm{a}}_{h}$) evolve in the linear regime, whereas the velocity and acceleration fluctuations ($\boldsymbol{\mathrm{v}}_{hp}^{i}$ and $\boldsymbol{\mathrm{a}}_{hp}^{i}$) evolve in the nonlinear regime (see Figs. \ref{fig:5} and \ref{fig:6}). In this paper, the mean velocity $\boldsymbol{\mathrm{v}}_{h}$ and acceleration $\boldsymbol{\mathrm{a}}_{h}$ refer to the mean velocity and acceleration of all particles in the same halo, i.e., the velocity and acceleration of that halo (Eq .\eqref{ZEqnNum680194}). Fluctuations ($\boldsymbol{\mathrm{v}}_{hp}^{i}$ and $\boldsymbol{\mathrm{a}}_{hp}^{i}$) refer to the velocity and acceleration of halo particles that fluctuate around their mean values $\boldsymbol{\mathrm{v}}_{h}$ and $\boldsymbol{\mathrm{a}}_{h}$ (Eq. \eqref{ZEqnNum576014}).

Figure \ref{fig:5} plots the time variation of the RMS acceleration ($\sqrt{3} \times$ standard deviation of the distributions in Fig. \ref{fig:4}), that is, the root mean square acceleration for halo particles (blue), out-of-halo particles (black), haloes (green), and acceleration fluctuations (red), where the factor $\sqrt{3}$ is for the magnitude of the acceleration vector in a 3D space. Halo acceleration $a_{h}$ is the mean acceleration of all particles in the same halo (Eq. \eqref{ZEqnNum680194}). All RMS accelerations decrease with time but evolve differently. The standard deviation of the halo acceleration $a_h$ and the out-of-halo particle acceleration $a_{op}$ evolves as $\propto a^{-1/2} \propto t^{-1/3}$ (the linear regime due to inter-halo interactions on large scales). The standard deviation of acceleration fluctuations $a_c=std(a_{hp}^i)$ evolves approximately as $a_c\propto a^{-3/4 }\propto t^{-1/2}$ in the nonlinear regime due to intra-halo interactions on small scales. The RMS acceleration for all particles $a_p$ can be approximated as a weighted average of out-of-halo particle acceleration $a_{op}$ and halo particle acceleration $a_{hp}$ (Eq. \eqref{eq:25}). It does increase with time due to the increasing number of halo particles. 

On large scales, haloes and out-of-halo particles have similar accelerations that are much smaller than the acceleration of halo particles due to the large distance between different haloes and between out-of-halo particles ($\approx 10^{-12}$m/s$^2$, green and black lines). At \textit{z}=0, the critical acceleration of fluctuations $a_c$ matches the empirical acceleration of BTFR $a_{0} \approx 10^{-10}$ m/s$^2$. This suggests a potential relationship between the critical acceleration $a_c$ (the RMS of acceleration fluctuations) and the empirical acceleration $a_0$.

Similarly, Figure \ref{fig:6} plots the time variation of the velocity variance (or specific kinetic energy). The halo velocity $v_{h}$ is the mean velocity of all particles in the same halo (Eq. \eqref{ZEqnNum680194}). The dispersion of the halo velocity $v_{h}$ matches the dispersion of the out-of-halo particle velocity $ v_{op}$. Both velocity dispersions have a linear scaling $\propto a$ that manifests the linear regime on large scales. Therefore, haloes behave like macro-particles on large scales with an evolution of acceleration and velocity similar to that of out-of-halo particles (see Figs. \ref{fig:5} and \ref{fig:6}). The velocity dispersions of haloes and out-of-halo particles evolve as ($\propto a \propto t^{2/3}$) due to inter-halo interactions on large scales that are in the linear regime. The dispersion of velocity fluctuations evolves as $\propto a^{{3/2}}\propto t$ due to the intra-halo interactions on small scales in the nonlinear regime. 

\section{Distributions of dark matter velocity}
\label{sec:4}
Since acceleration and velocity fluctuations are tightly connected in the self-gravitating collisionless dark matter, this section focuses on the velocity distribution in dark matter, which is also required to understand the origin of empirical acceleration $a_0$ in Section \ref{sec:8}. 

The velocity distributions of dark matter particles can be studied by N-body simulations and by the theory of maximum entropy \citep{Xu:2023-Maximum-entropy-distributions}. The principle of maximum entropy requires a velocity distribution with the highest entropy and the least prior information \citep{Jaynes:1957-Information-Theory-and-Statist-I,Jaynes:1957-Information-Theory-and-Statist-II}. In the kinetic theory of gases, the maximum entropy distribution of the molecule's velocity follows the well-known Boltzmann statistics,
\begin{equation} 
\label{ZEqnNum8641322} 
f\left(v\right)=\frac{1}{\sqrt{2\pi}v_{0}} \exp\left(-\frac{v^2}{2v_0^2}\right),         
\end{equation} 
where $v_0^2$ is the dispersion or the scale of velocity fluctuation. The Maxwell-Boltzmann distribution of particle speed reads  
\begin{equation} 
\label{ZEqnNum8641322-2} 
Z\left(v\right)=-2v\frac{\partial f}{\partial v} = \sqrt{\frac{2}{\pi}}\frac{v^2}{v_0^3}\exp\left(-\frac{v^2}{2v_0^2}\right).         
\end{equation}
For systems involving short-range interactions, the maximum entropy distribution of the velocity is Gaussian (Eq. \eqref{ZEqnNum8641322}). 

\begin{figure}
\includegraphics*[width=\columnwidth]{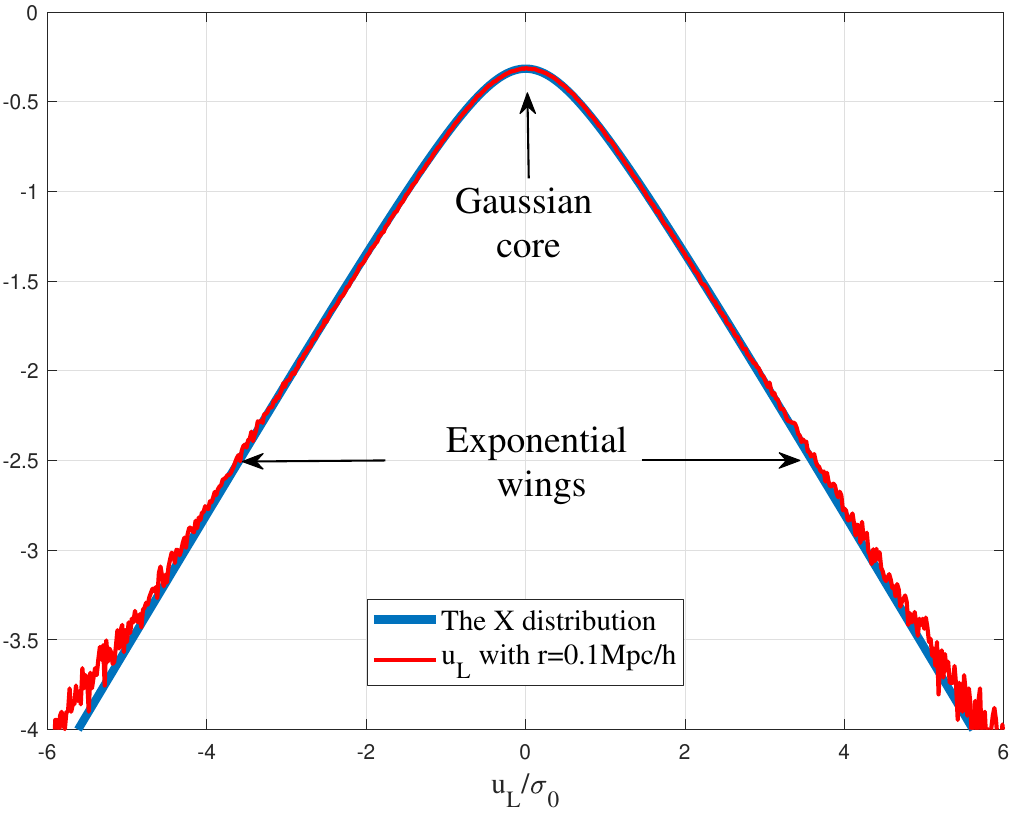}
\caption{The \textbf{\textit{X}} distribution with a unit variance compared with the velocity distribution from \textit{N}-body simulation ($u_{L} $ normalized by $\sigma _{0} $). The vertical axis is in the logarithmic scale (log$_{10}$). The \textbf{\textit{X}} distribution with $\alpha =1.33$ and $v_{0}^{2} ={1/3} \sigma _{0}^{2} $ matches the velocity distribution on a small scale $r$, where all pairs of particles are likely from the same halo. The Gaussian core ($u_L<v_0$) and exponential wings ($u_L>v_0$) can be clearly identified.}
\label{fig:2}
\end{figure}

For self-gravitating collisionless dark matter involving long-range gravity, the maximum entropy distribution of particle velocity (denoted as the \textbf{\textit{X}}) can also be analytically derived \citep{Xu:2023-Maximum-entropy-distributions}. This can be achieved by applying the virial theorem for mechanical equilibrium and the maximum entropy principle for statistical equilibrium. Here, we briefly present the main results. The maximum entropy distribution of particle velocity (\textbf{\textit{X}} distribution) reads \citep{Xu:2023-Maximum-entropy-distributions}
\begin{equation} 
\label{ZEqnNum864132} 
X\left(v\right)=\frac{1}{2\alpha v_{0} } \frac{e^{-\sqrt{\alpha ^{2} +\left({v/v_{0} } \right)^{2} } } }{K_{1} \left(\alpha \right)},        \end{equation} 
where $K_{y} \left(x\right)$ is a modified Bessel function of the \textit{second kind}. Similarly to Eq. \eqref{ZEqnNum8641322}, $v_{0}$ is a velocity scale (the scale of velocity fluctuation). Here, $\alphaup$ is a shape parameter that dominates the shape of \textbf{\textit{X}} distribution. The \textbf{\textit{X}} distribution approaches a double-sided Laplace (exponential) distribution with $\alpha \to 0$ and a Gaussian distribution with $\alpha \to \infty$, respectively. The second-order moment (variance) of the $X$ distribution reads: 
\begin{equation} 
\label{ZEqnNum108881} 
Var(v)=\alpha \frac{K_{2} \left(\alpha \right)}{K_{1} \left(\alpha \right)} v_{0}^{2} =\sigma _{0}^{2}.    
\end{equation} 
Similarly to the Maxwell-Boltzmann distribution in Eq. \eqref{ZEqnNum8641322-2}, the distribution of particle speed $Z(v)$ reads
\begin{equation} 
\label{ZEqnNum864132-2} 
Z\left(v\right)=-2v\frac{\partial X}{\partial v} = \frac{1}{\alpha K_{1} \left(\alpha \right)} \cdot \frac{v^{2} }{v_{0}^{3} } \cdot \frac{e^{-\sqrt{\alpha ^{2} +\left({v/v_{0} } \right)^{2} } } }{\sqrt{\alpha ^{2} +\left({v/v_{0} } \right)^{2} } }.         
\end{equation}

\begin{figure}
\includegraphics*[width=\columnwidth]{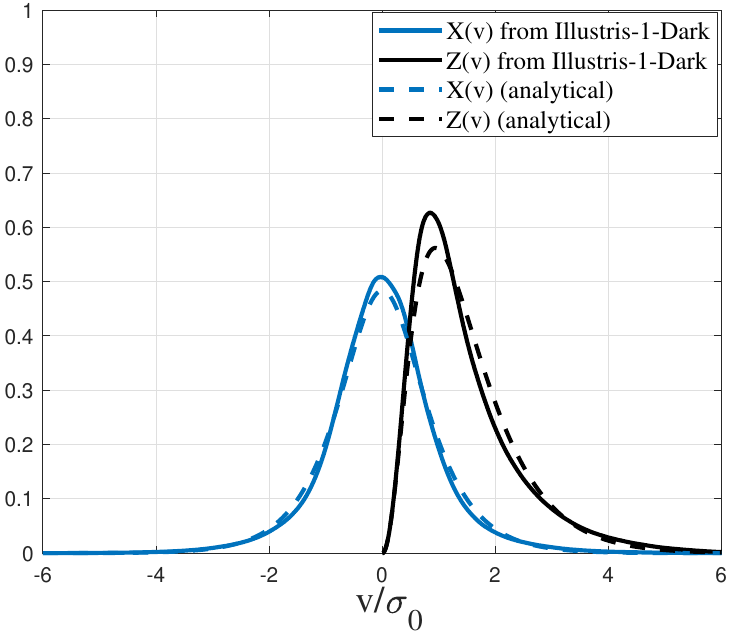}
\caption{The \textbf{\textit{X}} distribution for particle velocity and \textbf{\textit{Z}} distribution for particle speed from Illustris-1-dark simulation. Velocity is normalized by the standard deviation $\sigma _{0}$ in Eq. \eqref{ZEqnNum108881}. The shape parameter $\alpha =1.33$ and the velocity scale $v_{0}^{2} ={1/3} \sigma _{0}^{2}$. Dashed lines present the analytical model of velocity and speed distributions from Eqs. \eqref{ZEqnNum864132} and \eqref{ZEqnNum864132-2}.}
\label{fig:2-2}
\end{figure}

Figure \ref{fig:2} presents the comparison of \textbf{\textit{X}} distribution with the N-body simulation (SCDM). For all pairs of particles with a given separation $r$, the longitudinal velocity $u_{L} $ is calculated as the projection of the particle velocity $\boldsymbol{\mathrm{u}}$ along the vector of separation $\boldsymbol{\mathrm{r}}$, that is, $u_{L} =\boldsymbol{\mathrm{u}}\cdot \boldsymbol{\mathrm{r}}$. In this graph, the parameters $\alpha =1.33$, $v_{0}^{2} ={1/3} \sigma _{0}^{2}$ (from Eq. \eqref{ZEqnNum108881}), and the velocity variance $\sigma _{0}^{2}$ = var$\left(u_{L} \right)$ for all pairs of particles on a small scale of $r\equiv \left|\boldsymbol{\mathrm{r}}\right|=0.1{Mpc/h}$ at $z=0$ \citep{Xu:2023-Maximum-entropy-distributions,Xu:2024-On-the-statistical-theory-of-self-gravitating-scale_redshift}. 
For intermediate $\alphaup \sim 1$, the \textbf{\textit{X}} distribution naturally exhibits a Gaussian core at low velocity $v \ll v_{0}$ and exponential wings at high velocity $v \gg v_{0}$, as shown in Fig. \ref{fig:2}. Here, we find $v_0$, the scale of velocity fluctuation, that delineates two regimes. In this case, the \textbf{\textit{X}} distribution in Eq. \eqref{ZEqnNum864132} can be expressed separately, that is, the Gaussian core for small velocity and the exponential wings for large velocity: 
\begin{equation}
\begin{split}
&X\left(v\right)=\frac{e^{-\alpha } }{2\alpha v_{0} K_{1} \left(\alpha \right)} \exp \left(-\frac{v^{2} }{2\alpha v_{0}^{2} } \right) \quad   \textrm{for} \quad \left|v\right|\ll v_{0},\\
&X\left(v\right)=\frac{1}{2\alpha v_{0} K_{1} \left(\alpha \right)} \exp \left(-\frac{v}{v_{0} } \right) \quad \textrm{for} \quad \left|v\right|\gg v_{0}.  
\end{split}
\label{ZEqnNum174426}
\end{equation}
This is also consistent with other N-body simulation results \citep{Cooray:2002-Halo-models-of-large-scale-str}.

Figure \ref{fig:2-2} presents a similar comparison with the Illustris-1-Dark simulation \citep{NELSON:2015-The-illustris-simulation}. We first identify all haloes and halo particles. Solid lines present the velocity and speed distributions of all halo particles, while dashed lines plot the \textbf{\textit{X}} and \textbf{\textit{Z}} distributions from Eqs. \eqref{ZEqnNum864132} and \eqref{ZEqnNum864132-2}. Compared to the better agreement in Fig. \ref{fig:2}, the small discrepancy in Fig. \ref{fig:2-2} reflects the fact that some haloes may not be fully virialized. 

Comparing the \textbf{\textit{X}} distribution with the Gaussian distribution $f(v)$ in Eq. \eqref{ZEqnNum8641322}, the effect of long-range gravity on velocity distributions can be clearly demonstrated. For a small velocity $v \ll v_0$ of particles in the halo core region or small virialized haloes, the distribution is approximately Gaussian. For dark matter particles in the outer region of the haloes with a high velocity $v \gg v_0$, the velocity distribution deviates from Gaussian and approaches an exponential distribution, reflecting the effect of long-range inter-halo gravity on particles with high velocity.

Next, let us look at the specific kinetic energy of dark matter particles (energy per unit mass) that will provide more insight into the effect of long-range gravity. In Newtonian mechanics, the specific kinetic energy $\varepsilon_K(v)=v^2/2$. For self-gravitating collisionless dark matter, the specific kinetic energy $\varepsilon_K(v)$ for all particles with a given speed $v$ can be obtained from the maximum entropy distribution \citep{Xu:2023-Maximum-entropy-distributions},  
\begin{equation}
\label{ZEqnNum273068} 
\varepsilon_K \left(v\right)=-\frac{3}{2}\frac{X\left(v\right)v^2}{{v\partial X/\partial v} }=\frac{3X(v)v^2}{Z(v)}, 
\end{equation} 
where $X(v)v^2dv$ is the total kinetic energy for all particles with a given speed $v$, while $v\partial X/\partial v dv$ is the total number of particles with that given speed $v$ (e.g. Eq. \eqref{ZEqnNum864132-2}). Substitution of Eqs. \eqref{ZEqnNum864132} and \eqref{ZEqnNum864132-2} into Eq. \eqref{ZEqnNum273068} leads to the effective kinetic energy (Specific) as a function of particle speed $v$,
\begin{equation} 
\label{ZEqnNum520740} 
\varepsilon_K\left(v\right)=\frac{3}{2} v_{0}^{2} \sqrt{\alpha ^{2} +\left(\frac{v}{v_{0} } \right)^{2} }. 
\end{equation} 
Note that $\varepsilon_K$ represents the effective kinetic energy. 

In N-body simulation, we can identify all particles with a given speed $v$ at any moment with instantaneous energy of $\varepsilon_i(v)=v^2/2+\phi_i$, where $\phi_i$ is the potential energy of that particle at that moment. The average total energy of all particles with the same speed $v$ can be written as
\begin{equation} 
\label{ZEqnNum520740-2} 
\langle \varepsilon_i(v)\rangle = \varepsilon_K + \varepsilon_P \quad \textrm{and} \quad 2\varepsilon_K = n \varepsilon_P,
\end{equation} 
where the effective kinetic energy $\varepsilon_K$ and potential energy $\varepsilon_P$ obey the virial theorem and $n$ is the potential exponent. Particles with the same speed $v$ can be in haloes of different sizes with different instantaneous potential $\phi_i$. The virial theorem is valid in the sense of averaging over an ensemble of all particles with the same speed $v$ or averaging over the time trajectory of a given particle. Therefore, the effective particle kinetic energy $\varepsilon_K$ reflects the averaged particle kinetic energy for all particles with the same speed. 

For self-gravitating dark matter, the particle kinetic energy $\varepsilon_K \left(v\right)$ has two regimes from Eq. \eqref{ZEqnNum520740} that follow a linear scaling and a parabolic scaling, respectively,
\begin{equation}
\begin{split}
&\varepsilon_K \left(v\right)\approx \frac{3}{2} v_{0} v \quad \textrm{for} \quad v\gg v_{0},\\ 
&\varepsilon_K \left(v\right)\approx \underbrace{\frac{3}{2} \alpha v_{0}^{2}}_{1} +\underbrace{\frac{3v^{2} }{4\alpha }}_{2} \quad \textrm{for} \quad v\ll v_0.
\end{split}
\label{ZEqnNum201633}
\end{equation}
For particles with a small velocity $v\ll v_0$, term 1 is the contribution from the halo velocity $v_h$, i.e., the mean velocity of all particles in the same halo. High-speed particles are usually in the outskirts of haloes with extremely small acceleration. Due to the long-range nature of gravity, their dynamics are dominated by both inter- and intra-halo interactions. The superposition of all intra- and inter-halo interactions leads to the linear scaling for kinetic energy $\varepsilon _{K} \left(v\right)\propto v_0v$, which emerges from the maximum entropy principle. This scaling can also be confirmed by N-body simulations \citep{Xu:2023-Maximum-entropy-distributions}. On the contrary, the dynamics of low-speed particles in the halo core region are dominated only by the intra-halo interaction with Newtonian behavior, that is, $\varepsilon _{K} \left(v\right)\propto v^{2}$.  

In summary, the velocity distributions of dark matter involve a typical velocity scale $v_0$ (the scale of fluctuation). Depending on the particle velocity $v$, the effective kinetic energy $\varepsilon_{K}$ follows $\varepsilon _{K} \left(v\right)\propto v^{2}$ for low speed ($v\ll v_{0} $), that is, a standard Newtonian behavior. However, $\varepsilon _{K} \left(v\right)\propto v_{0} v$ for high-speed ($v\gg v_{0}$) is a unique "non-Newtonian" feature due to long-range gravity in dark matter. These results will be used to explain the non-Newtonian behavior and the empirical MOND acceleration $a_0$ in Section \ref{sec:8}. 

\begin{figure}
\includegraphics*[width=\columnwidth]{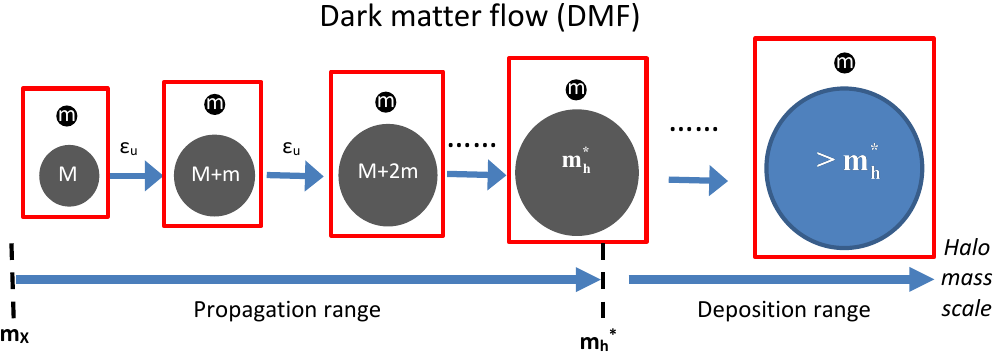}
\caption{Schematic plot of the mass and energy cascade in dark matter flow. Hierarchical structure formation proceeds via halo merging to give rise to larger haloes. Haloes merge with smaller haloes to give rise to larger haloes and produce a continuous mass/energy flux to larger mass scales, that is, a continuous inverse mass and kinetic energy cascade in the halo mass space. While the potential energy is directly cascaded from large to small scales. Scale-independent rate of mass cascade ($\varepsilon_m$ in Eq. \eqref{ZEqnNum986319}) and rate of (kinetic) energy cascade ($\varepsilon_{u}$ in Eq. \eqref{ZEqnNum98631913}) are expected in a certain range of mass scales (propagation range). The mass/energy cascaded from small scales is consumed to grow haloes on scales greater than $m_h^*$ (deposition range). Figure \ref{fig:8} presents a typical two-body merging to calculate the rate of the energy cascade $\varepsilon_{u}$. Table \ref{tab:2} presents the relevant physical quantities on the smallest scale ($m_X$) and the largest scale ($m_h^*$) of the propagation range. More details can be found in previous work \citep{Xu:2023-Universal-scaling-laws-and-density-slope,Xu:2023-Dark-matter-halo-mass-functions-and,Xu:2022-Postulating-dark-matter-partic,Xu:2021-Inverse-mass-cascade-mass-function} and the applications for universal halo density profile, halo mass function, and dark matter particle properties.} 
\label{fig:S2}
\end{figure}

\section{The energy cascade in dark matter}
\label{sec:6-1}
The acceleration and velocity fluctuations are not independent of each other. In fact, two fluctuations are intrinsically connected through an energy cascade process in dark matter, which has been extensively discussed in previous work. Halo mass functions, density profiles, and universal scaling laws can be systematically derived based on the concept of energy cascade across haloes of different sizes \citep{Xu:2023-Universal-scaling-laws-and-density-slope,Xu:2023-Dark-matter-halo-mass-functions-and, Xu:2021-Inverse-mass-cascade-mass-function}, where the readers can find more details. Here, we provide a brief overview, as shown in Fig. \ref{fig:S2}, followed by illustrations of the concept from the results of N-body simulations. 

First, long-range gravity requires the formation of different sizes of haloes to maximize the entropy of the system \citep{Xu:2023-Maximum-entropy-distributions}. These localized overdense halo structures (building blocks in dark matter flow) facilitate the formation of hierarchical structures via halo merging to give rise to larger and larger structures (see Fig. \ref{fig:S2}). Due to hierarchical structure formation, haloes of mass $M$ merge with a smaller halo of mass $m$ to produce a mass/energy flux into the scale of $M+m$ in the halo mass space. Continuous halo structure merging facilitates a continuous mass/energy cascade (or flux) from small to large scales to form larger and larger structures. This halo-mediated cascade process, combined with acceleration fluctuations, improves our understanding of the empirical acceleration $a_0$ in BTFR.

Second, a constant rate of the energy cascade $\varepsilon_{u}$ is expected in certain ranges of halo mass scales, that is, the propagation range 
, where $\varepsilon_{u}$ is independent of the mass scale $m_h$ (Fig. \ref{fig:S1-1-6}). If this is not the case, there would be a net accumulation of energy on some intermediate-mass scale below $m_{h}^{*}$. We exclude this possibility because we require the statistical structures of the haloes to be self-similar and scale-free for haloes smaller than $m_{h}^{*}$. This leads to scale-independent cascade rates up to a critical mass $m_h^*$. Mass/energy cascaded from small scales is consumed to grow haloes beyond the propagation range, i.e., a deposition range. These fundamental concepts can be rigorously formulated and demonstrated by N-body simulations, along with the value of the rate of cascade $\varepsilon_u$. A similar analysis can also be extended to the mass and energy flow in galaxy bulges for relevant scaling laws and dynamic evolution of galaxies and supermassive black holes \citep{Xu:2024-Cosmic-quenching-and-scaling-laws}. 

\begin{figure}
\includegraphics*[width=\columnwidth]{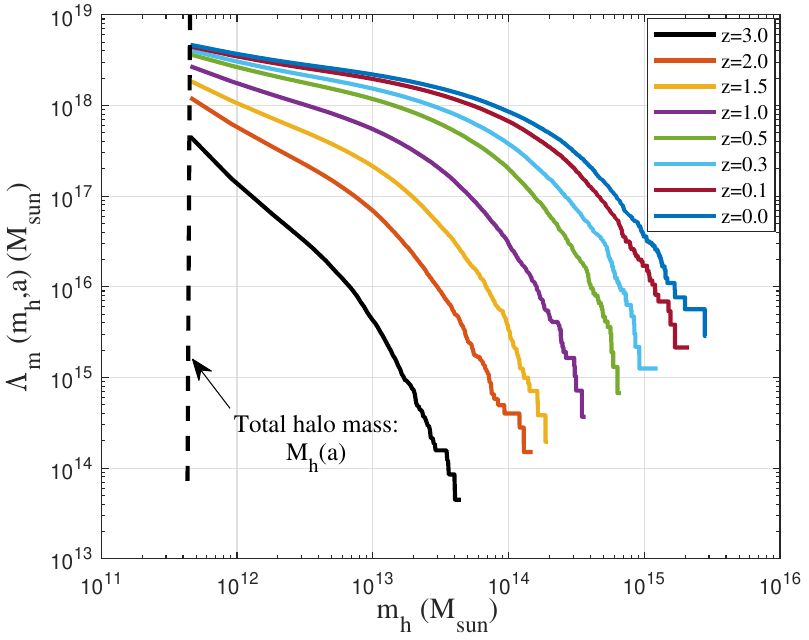}
\caption{The variation of cumulative mass function $\Lambda_m(m_h,a)$ with halo mass scale $m_h$ at different redshifts $z$. The total mass $M_h(a)$ in all haloes of all sizes is calculated with $m_h\rightarrow 0$ (Eq. \eqref{ZEqnNum98631129}), i.e. $M_h(a)=\Lambda_m(m_h=0,a)$ that increases with time.} 
\label{fig:S1-1}
\end{figure}

\begin{figure}
\includegraphics*[width=\columnwidth]{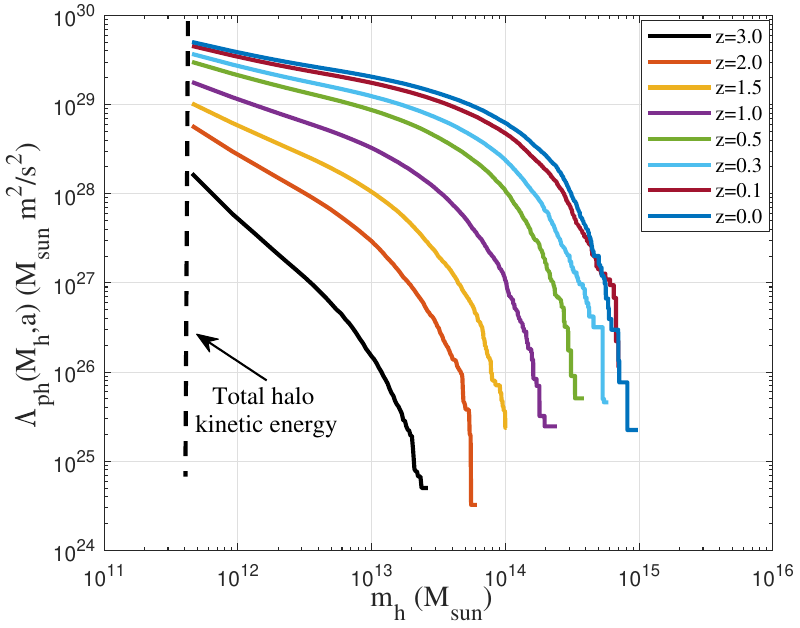}
\caption{The variation of cumulative halo kinetic energy $\Lambda_{ph}(m_h, a)$ with halo mass scale $m_h$ at different redshifts $z$, i.e., the kinetic energy from the mean velocity $v_h$ (the motion of entire halo) due to inter-halo interactions on large scales in the linear regime.}
\label{fig:S1-1-1}
\end{figure}

\begin{figure}
\includegraphics*[width=\columnwidth]{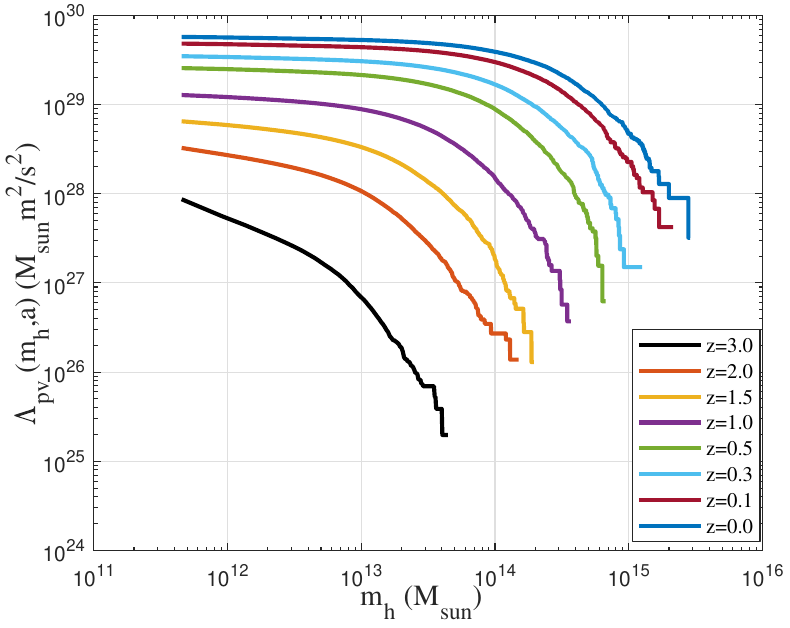}
\caption{The variation of cumulative halo kinetic energy $\Lambda_{pv}(m_h, a)$ with halo mass scale $m_h$ at different redshifts $z$, i.e. the kinetic energy from the velocity fluctuation $v_{hp}^i$ (Eq. \eqref{ZEqnNum576014}) due to intra-halo interactions on small scales in the nonlinear regime.}
\label{fig:S1-1-2}
\end{figure}

To quantify the mass and energy cascade, we first introduce a cumulative mass function $\Lambda_m(m_h, a)$ that represents the total mass in all haloes greater than a mass scale $m_h$,
\begin{equation} 
\label{ZEqnNum98631129} 
\Lambda_m(m_h,a) = \int _{m_{h}}^{\infty } M_{h} \left(a\right)f_{M} \left(m,m_{h}^{*} \right) dm,  
\end{equation} 
where $a$ is the scale factor and $M_{h}$ is the total mass in all haloes of all sizes. Here, $f_{M}(m_h,m_{h}^{*})$ is the halo mass function that gives the probability distribution of the total mass $M_h$ in haloes of different mass $m_h$. Figure \ref{fig:S1-1} plots the variation of the cumulative mass function $\Lambda_m(m_h, a)$ with halo mass $m_h$ and redshifts $z$. The total halo mass $M_h$ can be obtained by setting $m_h\rightarrow 0$ in Eq. \eqref{ZEqnNum98631129}, i.e. $M_h(a)=\Lambda_m(m_h\rightarrow 0,a)$ that increases with time. 

The mass flux ($\Pi_m$) across haloes of different sizes reads
\begin{equation} 
\label{ZEqnNum986318} 
\begin{split}
&\Pi _{m} \left(m_{h} ,a\right) =-\int _{m_{h} }^{\infty }\frac{\partial }{\partial t} \left[M_{h} \left(a\right)f_{M} \left(m,m_{h}^{*} \right)\right] dm, \\
&=-\frac{\partial }{\partial t} \left[\int _{m_{h} }^{\infty } M_{h} \left(a\right)f_{M} \left(m,m_{h}^{*} \right) dm \right]=-\frac{\partial \Lambda_m}{\partial t}.
\end{split}
\end{equation} 
The change of total mass $\Lambda_m$ above scale $m_h$ equals the mass cascaded from scales below scale $m_h$. As shown in the schematic plot in Fig. \ref{fig:S2}, all masses cascaded from the smallest scale are propagated through the propagation range and are consumed mainly to grow haloes greater than $m_h^*$. The scale-independent mass flux in the propagation range reads
\begin{equation} 
\label{ZEqnNum986319} 
\begin{split}
&\varepsilon_m(a) \equiv \Pi _{m}(m_h,a) \quad \textrm{for}\quad m_h<m_h^*.
\end{split}
\end{equation}

Next, velocity $v_{hp}$ of every halo particle can be decomposed into mean velocity $v_h$ and velocity fluctuation $v_{hp}^i$ according to Eq. \eqref{ZEqnNum576014}. Similarly, the kinetic energy of each halo particle can be decomposed, that is, the halo kinetic energy from the mean velocity $K_{ph}=v_h^2/2$ and the virial kinetic energy $K_{pv}=(v_{hp}^i)^2/2$ from the velocity fluctuation $v_{hp}^i$. 

Similarly to the cumulative mass function $\Lambda_m$ in Eq. \eqref{ZEqnNum98631129}, the cumulative kinetic energies ($\Lambda_{ph}$ and $\Lambda_{pv}$) can be introduced to represent the total kinetic energies $K_{ph}$ and $K_{pv}$ in all haloes greater than the mass scale $m_h$,
\begin{equation} 
\label{ZEqnNum986311299} 
\begin{split}
&\Lambda_{ph}(m_h,a) = \int _{m_{h}}^{\infty } M_{h} \left(a\right)f_{M} \left(m,m_{h}^{*} \right) K_{ph} dm,  \\
&\Lambda_{pv}(m_h,a) = \int _{m_{h}}^{\infty } M_{h} \left(a\right)f_{M} \left(m,m_{h}^{*} \right) K_{pv} dm, \\
\end{split}
\end{equation} 

Using Virgo simulations (SCDM), Figs. \ref{fig:S1-1-1} and \ref{fig:S1-1-2} present the variation of the cumulative halo kinetic energy $\Lambda_{ph}$ and the virial kinetic energy $\Lambda_{pv}$ with halo mass scale $m_h$ at different redshifts $z$. 

\begin{figure}
\includegraphics*[width=\columnwidth]{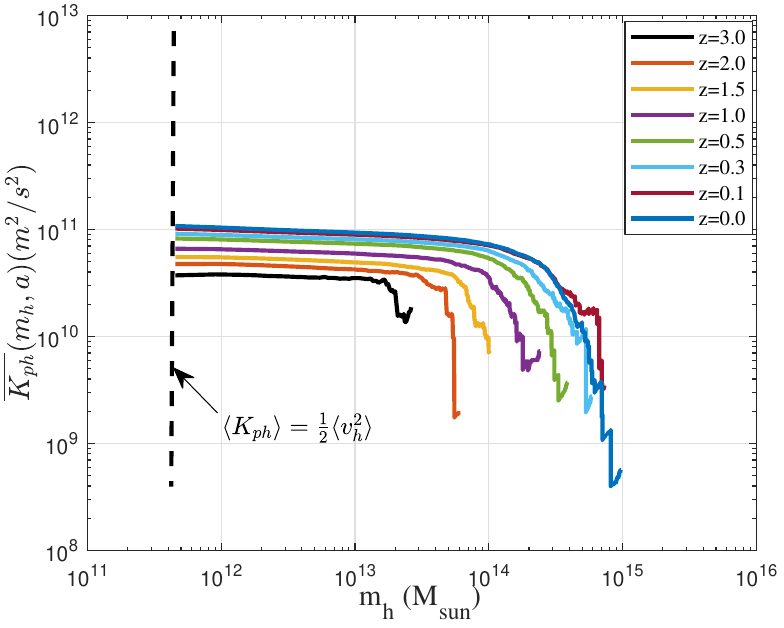}
\caption{The variation of mean halo kinetic energy $\overline {K_{ph}}$ with halo mass scale $m_h$ at different redshifts $z$. The mean halo kinetic energy $\langle K_{ph} \rangle$ in all haloes of all sizes is denoted as the dashed line with $m_h\rightarrow 0$, i.e. $\langle K_{ph} \rangle=\overline {K_{ph}}(m_h\rightarrow 0, a)$ that is also related to the velocity dispersion $\langle v_h^2\rangle$ in Fig. \ref{fig:6}. The time evolution of $\langle {K_{ph}}\rangle \propto t^{2/3}$ is shown in Fig. \ref{fig:S1-1-7}.} 
\label{fig:S1-1-4}
\end{figure}

\begin{figure}
\includegraphics*[width=\columnwidth]{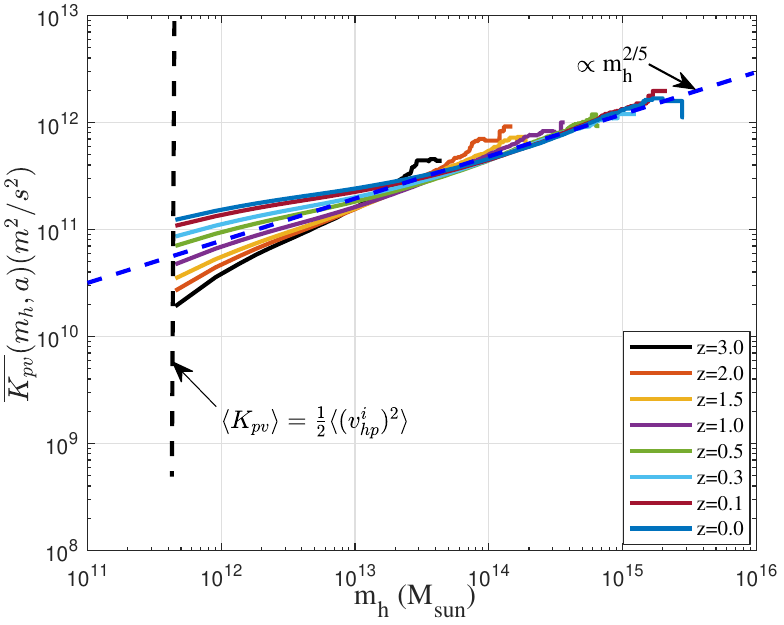}
\caption{The variation of mean virial kinetic energy $\overline {K_{pv}}$ with halo mass scale $m_h$ at different redshifts $z$. The mean virial kinetic energy $\langle K_{pv} \rangle$ in all haloes of all sizes is denoted as the dashed line with $m_h\rightarrow 0$. Here, $\langle K_{pv} \rangle=\overline {K_{pv}}(m_h\rightarrow 0,a)$ is also related to the velocity dispersion $\langle (v_{hp}^i)^2\rangle$ in Fig. \ref{fig:6}. This figure will be used to compute the rate of the energy cascade $\varepsilon_u$ in Fig. \ref{fig:S1-1-6}. The time evolution of virial kinetic energy $K_{pv}$ (the non-linear regime with $\langle {K_{pv}}\rangle\propto t^{1}$ is also shown in Fig. \ref{fig:S1-1-7}). The virial kinetic energy follows a 2/5 scaling for large haloes, where $\overline {K_{pv}}\propto m_h^{2/5}$.} 
\label{fig:S1-1-5}
\end{figure}

Next, using the cumulative kinetic energy and the cumulative mass $\Lambda_m$, we can calculate the mean (specific) halo kinetic energy $\overline {K_{ph}}$ and the mean (specific) virial kinetic energy $\overline {K_{pv}}$ (energy per unit mass) in all haloes above any mass scale $m_h$, that is,
\begin{equation} 
\label{ZEqnNum98631129911} 
\overline {K_{ph}}=\frac{\Lambda_{ph}}{\Lambda_{m}} \quad \textrm{and} \quad \overline {K_{pv}}=\frac{\Lambda_{pv}}{\Lambda_{m}}.
\end{equation}
Figures \ref{fig:S1-1-4} and \ref{fig:S1-1-5} plot the variation of the halo kinetic energy $\overline {K_{ph}}$ and the virial kinetic energy $\overline {K_{pv}}$ with the mass scale $m_h$. The halo kinetic energy is relatively independent of $m_h$. The virial kinetic energy increases with the mass of the halo $m_h$ as $\overline {K_{pv}}\propto m_h^{2/5}$ for larger haloes with mass $m_h\rightarrow \infty$ \citep{Xu:2022-Postulating-dark-matter-partic}. Both kinetic energies increase with time but with different scalings because of the distinct nature of intra- and inter-halo interactions in the linear and nonlinear regimes. 

The energy cascade is often associated with the kinetic energy contained in the random velocity fluctuations or $\overline {K_{pv}}$. Next, we will focus on the cascade of the specific virial kinetic energy $K_{pv}$, which is due to the nonlinear interactions and fluctuations on small scales. The rate of cascade for (specific) virial kinetic energy $K_{pv}$ is defined as (similarly to the mass flux defined in Eq. \eqref{ZEqnNum986318})
\begin{equation} 
\label{ZEqnNum9863112991} 
\begin{split}
&\Pi_{pv}(m_h,a) =-\frac{\partial }{\partial t} \left( \overline {K_{pv}} \right) = -\frac{\partial }{\partial t} \left(\frac{\Lambda_{pv}}{\Lambda_m} \right) \\
&=-\frac{\partial }{\partial t}\int _{m_{h} }^{\infty } \frac{M_{h} \left(a\right)f_{M} \left(m,m_{h}^{*} \right) K_{pv}}{\int _{m_{h} }^{\infty } M_{h} \left(a\right)f_{M} \left(m,m_{h}^{*} \right) dm} dm,
\end{split}
\end{equation}
where $\overline {K_{pv}}$ is defined in Eq. \eqref{ZEqnNum98631129911}, i.e., the specific virial kinetic energy in all haloes greater than $m_h$. The inverse cascade (transfer) of the virial kinetic energy $K_{pv}$ from all scales below $m_h$ leads to the change of $K_{pv}$ on all scales above $m_h$, that is, the change of $\overline {K_{pv}}$ with time. Therefore, Eq. \eqref{ZEqnNum9863112991} describes the transfer of specific virial kinetic energy ($K_{pv}$) from all haloes below the scale $m_h$ to all haloes above the scale $m_h$ at a rate of $\Pi_{pv}$, i.e. the rate of energy cascade.

\begin{figure}
\includegraphics*[width=\columnwidth]{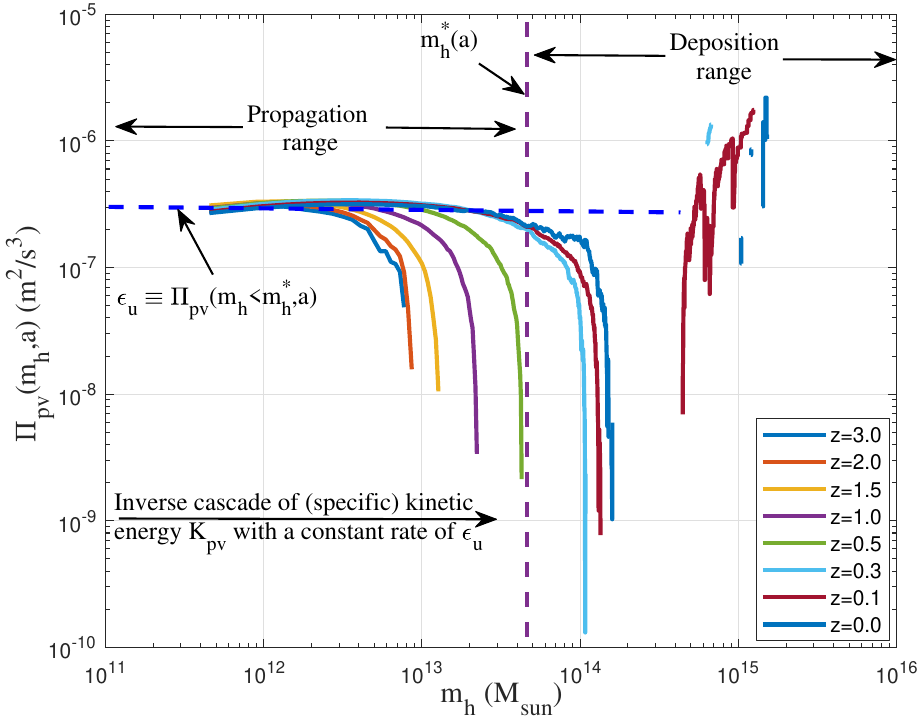}
\caption{The variation of the rate of energy cascade $\Pi_{pv}(m_h,a)$ (using Eq. \eqref{ZEqnNum9863112991} and data in Fig. \ref{fig:S1-1-5}) with halo mass scale $m_h$ at different redshifts $z$. A scale-independent constant rate of $\varepsilon_u$ can be clearly identified in the propagation range for an inverse cascade of virial kinetic energy ${K_{pv}}$ from small scales to larger scales. That rate is also relatively independent of time and is around -$3\times 10^{-7}m^2/s^3$ (also see Fig. \ref{fig:S1-1-7}). This is an important quantity that connects the velocity and acceleration fluctuations in Section \ref{sec:6}. The (purple) dashed line denotes the characteristic mass $m_h^*$ at $z=0$.} 
\label{fig:S1-1-6}
\end{figure}

Figure \ref{fig:S1-1-6} plots the variation of $\Pi_{pv}$ with the mass scale $m_h$ and the redshifts $z$. The mean (specific) virial kinetic energy $\overline {K_{pv}}$ in Fig. \ref{fig:S1-1-5} was used to calculate the cascade rate $\Pi_{pv}$ in this figure. The simulation confirms that the rate of the energy cascade $\Pi_{pv}$ is independent of the mass scale $m_h$ and the time $t$ in the propagation range $m_h<m_h^*$ in Fig. \ref{fig:S2} since the statistical structures of the haloes should be self-similar and scale-free for haloes smaller than $m_h^*$. Therefore, in the propagation range, we can write
\begin{equation} 
\label{ZEqnNum98631913} 
\begin{split}
\varepsilon_u \equiv \Pi _{pv}(m_h,a) \quad \textrm{for}\quad m_h<m_h^*,
\end{split}
\end{equation} 
where $\varepsilon_u$ is a very important constant that reflects the rate of energy cascade across haloes of different sizes. The physical meaning and impact of the scale-independent rate of cascade were further discussed for the structure and evolution of dark matter haloes \citep{Xu:2021-Inverse-mass-cascade-mass-function} and galaxies \citep{Xu:2024-Cosmic-quenching-and-scaling-laws}.

\begin{figure}
\includegraphics*[width=\columnwidth]{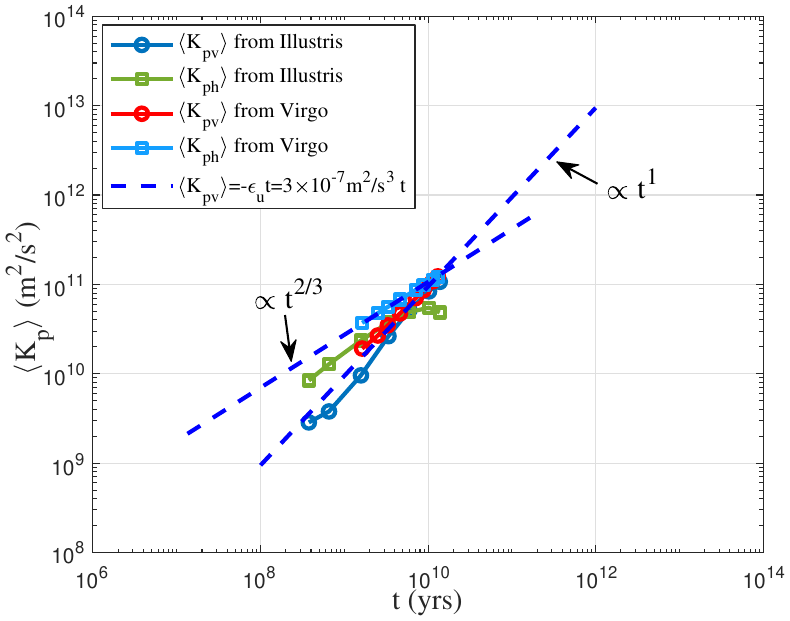}
\caption{The variation of the mean specific halo kinetic energy (energy per unit mass) $\langle K_{ph}\rangle$ and virial kinetic energy $\langle K_{pv}\rangle$ (unit: $m^2/s^2$) for all halo particles from Virgo and Illustris-1-Dark simulations. Simulations confirm a constant rate of the energy cascade $\varepsilon_u=-3\times 10^{-7}m^2/s^3$. Due to the long-range inter-halo interactions with particles from different haloes, $\langle K_{ph}\rangle$ increases with time in the early matter dominant universe (i.e., the linear regime with $\langle K_{ph}\rangle \propto t^{2/3}$). In the Illustris simulation, it slightly decreases at low redshift in the dark energy dominant universe due to the accelerated expansion. Due to intra-halo interactions with particles from the same halo, the virial kinetic energy $K_{pv}$ increases as $\langle K_{pv}\rangle\propto -\varepsilon_u t$ on small scales (the non-linear regime). Dark energy has smaller effects on $\langle K_{pv}\rangle$ because of bounded structures on small scales.} 
\label{fig:S1-1-7}
\end{figure}

Acceleration and velocity fluctuations are connected by $\varepsilon_u$ in Eq. \eqref{ZEqnNum219659}. With continuous injection of virial kinetic energy $K_{pv}$ at a constant rate of $\varepsilon_u$ on the smallest scale (mass $m_X$ in Fig. \ref{fig:S2}), we expect the total $K_{pv}$ in all haloes of all sizes to be proportional to time $t$, that is, $\langle K_{pv}\rangle\propto -\varepsilon_ut$, as shown in Fig. \ref{fig:S1-1-7}. Figure \ref{fig:S1-1-7} presents the evolution of the mean specific halo kinetic energy $\langle K_{ph}\rangle$ and the virial kinetic energy $\langle K_{pv}\rangle$ for all halo particles, i.e. the specific kinetic energy $\overline {K_{ph}}$ and $\overline {K_{pv}}$ in Figs. \ref{fig:S1-1-4} and \ref{fig:S1-1-5} with the limit $m_h\rightarrow 0$ (dashed lines). The data in Figs. \ref{fig:S1-1-4} and \ref{fig:S1-1-5} were used to plot the time evolution. A constant rate of the energy cascade $\varepsilon_u=-3\times 10^{-7}m^2/s^3$ can be confirmed as the proportional constant for $\langle K_{pv}\rangle$ (that is, $\langle K_{pv}\rangle\propto -\varepsilon_ut$), where $\varepsilon_u<0$ represents the inverse cascade from small to large scales). The halo kinetic energy $\langle K_{ph}\rangle \propto a\propto t^{2/3}$ is due to inter-halo interactions on large scales, which is in the linear regime. In the Illustris simulation, it levels off at low redshift $z$ because of the effect of dark energy. While on small scales or in the non-linear regime, the virial kinetic energy $\langle K_{pv}\rangle \propto -\varepsilon_u t$ is due to the inverse cascade of the kinetic energy. The virial kinetic energy is less affected by dark energy because of the bounded halo structure on small scales.

In summary, this section introduces the concept of energy cascade in dark matter. The kinetic energy is inversely cascaded from small to large scales at a constant rate $\varepsilon_u$. N-body simulations confirm a constant value of $\varepsilon_u\approx -3\times 10^{-7}m^2/s^3$ (Figs. \ref{fig:S1-1-6} and \ref{fig:S1-1-7}). The same value of this fundamental constant can also be obtained from Illustris simulations and discussed elsewhere \citep{Xu:2022-Postulating-dark-matter-partic, Xu:2023-Universal-scaling-laws-and-density-slope,Xu:2023-Dark-matter-halo-mass-functions-and, Xu:2021-Inverse-mass-cascade-mass-function}. In the next section, we will use these results to connect the acceleration and velocity fluctuations and identify the critical acceleration $a_c$ (the scale of fluctuation).

\section{The critical acceleration $a_c$ from energy cascade}
\label{sec:6}
\begin{figure}
\includegraphics*[width=\columnwidth]{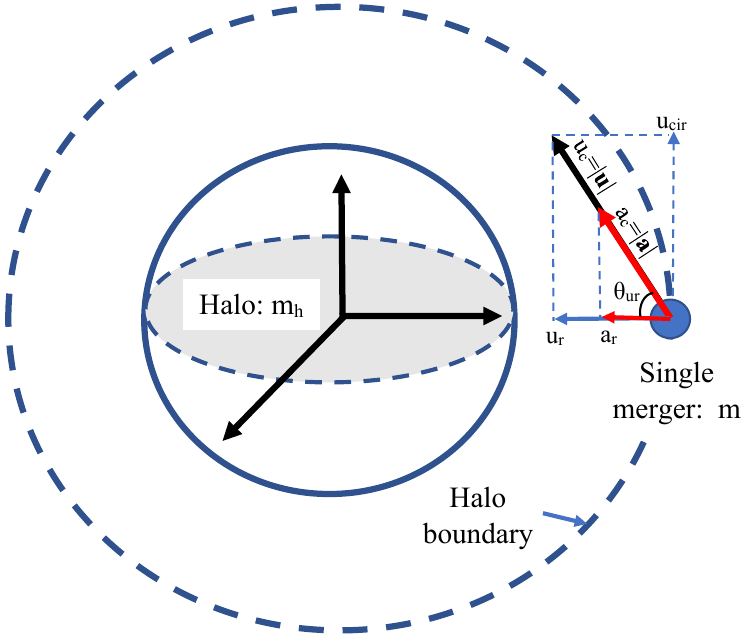}
\caption{The schematic plot of a typical two-body merging at the instantaneous moment of merging between a halo (mass: $m_{h}$) and a single merger (mass: \textit{m}). The single merger has a typical velocity $\boldsymbol{\mathrm{u}}$ (black) and acceleration $\boldsymbol{\mathrm{a}}$ (red). The dashed line is the boundary of that halo. The angle of incidence satisfies $\cot \left(\theta _{ur} \right)={1}/{(3\pi)}$. Due to the two-body merging, halo mass $m_h$ is cascaded to a larger mass scale at that instantaneous moment, i.e., a halo of mass $m_h$ moving into mass scale $m_h+m$ at that moment. In addition, the specific kinetic energy is simultaneously transferred to larger mass scale at a rate of $\varepsilon_u\propto \boldsymbol{\mathrm{a}} \cdot \boldsymbol{\mathrm{u}} \propto a_cu_c$ (Eq. \eqref{ZEqnNum219659}) at that moment.}
\label{fig:8}
\end{figure}

The energy cascade in dark matter is closely related to velocity and acceleration fluctuations, where the constant rate of the energy cascade $\varepsilon_u$ obtained in Fig. \ref{fig:S1-1-6} can be directly calculated from these fluctuations (Eq. \eqref{ZEqnNum219659}). To demonstrate this, consider an elementary merging between a halo and a single merger, which facilitates the continuous energy cascade across haloes of different scales. In a finite time interval $\Delta$t, the merging of hierarchical structures might involve multiple substructures merging into a single large structure. For an infinitesimal time interval \textit{dt}, that process should involve the merging of two and only two substructures (Fig. \ref{fig:8}) such that the two-body merging is the most elementary and frequent process for structures formation \citep{Xu:2021-A-non-radial-two-body-collapse}. The continuous two-body merging between the halo and single mergers facilitates the hierarchical structure formation and mass and energy cascade in Fig. \ref{fig:S2}. 

Now consider a typical two-body merging for a dark matter halo of mass $m_h$ (Fig. \ref{fig:8}). This halo involves velocity and acceleration fluctuations with a critical velocity $u_c\equiv\lvert\boldsymbol{\mathrm{u}}\rvert$ and a critical acceleration $a_c\equiv\lvert\boldsymbol{\mathrm{a}}\rvert$, representing the scale of velocity and acceleration fluctuations, respectively. These are also the RMS velocity fluctuation ($v_{hp}^i$) and the acceleration fluctuation ($a_{hp}^i$) for all halo particles in Eq. \eqref{ZEqnNum576014} and Figs. \ref{fig:5} and \ref{fig:6}. Figure \ref{fig:8} plots the instantaneous moment of a typical two-body merging when a single merger with a mass \textit{m}, a velocity vector $\boldsymbol{\mathrm{u}}$ due to the velocity fluctuations, and an acceleration vector $\boldsymbol{\mathrm{a}}$ due to the acceleration fluctuations, merging with a halo of mass $m_{h}$. The single merger is right on the halo boundary (dashed line) at the moment of merging. 
\newline 

\noindent i) We focus on the instantaneous moment of merging such that the single merger is right on the halo boundary. After this moment, the halo of mass $m_h$ will move to the next mass scale $m_h+m$, that is, the cascade of mass $m_h$ from the scale $m_h$ to larger mass scales $m_h+m$ (Fig. \ref{fig:S2}). Now, we focus on the energy cascade that is associated with the mass cascade. Figure \ref{fig:S1-1-6} describes the calculation of the rate of the energy cascade $\varepsilon_u$ based on the definition in Eq. \eqref{ZEqnNum9863112991}. This section provides an alternative way to calculate $\varepsilon_u$ based on the velocity and acceleration fluctuations; 
\newline 

\noindent ii) We focus on the cascade of the (specific) virial kinetic energy due to intra-halo interactions on the nonlinear small scales (Fig. \ref{fig:S1-1-6}). Only velocity and acceleration fluctuations are relevant (Eq. \eqref{ZEqnNum576014}), not the mean halo velocity and the mean halo acceleration. Therefore, both velocity $\boldsymbol{\mathrm{u}}$ and acceleration $\boldsymbol{\mathrm{a}}$ are the fluctuations in velocity and acceleration relative to the center of the halo. 
\newline

\begin{figure}
\includegraphics*[width=\columnwidth]{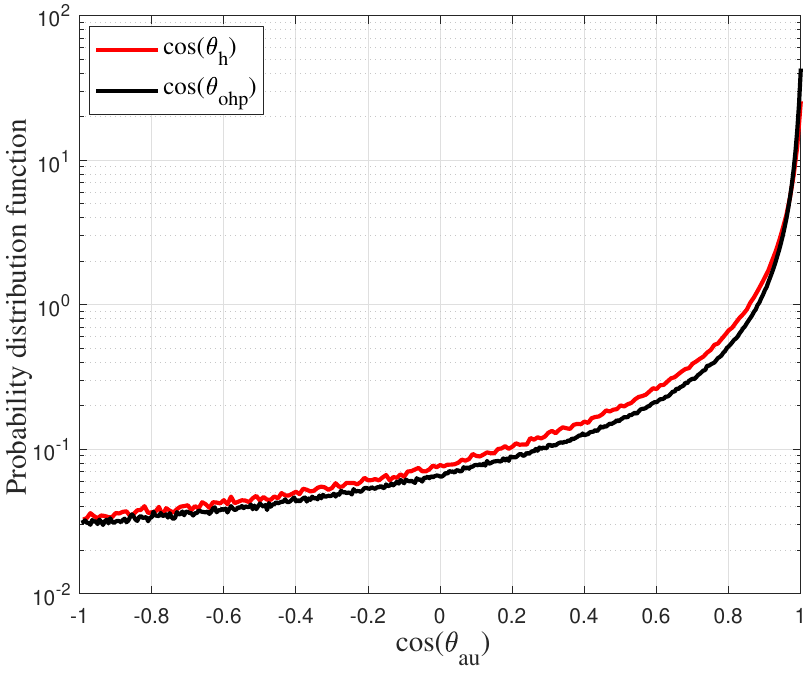}
\caption{The probability distribution of cos($\theta$) from Virgo N-body simulations for the angle $\theta$ between velocity and acceleration vectors. Here, $\theta_h$ represents the angle between halo velocity $\boldsymbol{\mathrm{v}}_{h}$ and halo acceleration $\boldsymbol{\mathrm{a}}_{h}$ in Eq. \eqref{ZEqnNum576014}, $\theta_{ohp}$ represents the angle between velocity and acceleration for "out-of-halo" particles. This plot shows the velocity and acceleration of out-of-halo particles and haloes aligned on large scales pointing in the same direction. Most haloes and out-of-halo particles have small angles with cos($\theta$)$\approx$ 1.} 
\label{fig:8-2}
\end{figure}

\noindent iii) Due to the Zeldovich approximation \citep{Zeldovich:1970-Gravitational-Instability-an} in the linear regime, a single merger, as an out-of-halo particle before merging, should have its velocity aligned with the direction of force or acceleration, that is, $\partial \boldsymbol{\mathrm{u}}/\partial t = H \boldsymbol{\mathrm{u}}/2$ on large scales. Therefore, at the halo boundary, the velocity vector $\boldsymbol{\mathrm{u}}$ and the acceleration vector $\boldsymbol{\mathrm{a}}$ of a single merger are more likely to align with each other, that is, $\boldsymbol{\mathrm{u}}$ and $\boldsymbol{\mathrm{a}}$ point in the same direction at the moment of merging, as shown in Figs. \ref{fig:8} and \ref{fig:8-2}. 
\newline \newline
\noindent iv) Due to the gravity of the halo to merge with, the single merger right on the boundary (dashed line) has a relative motion towards the center of the halo $u_{r} =u_c\cos (\theta _{ur})$ (radial velocity) and a radial acceleration with $a_{r} =a_c \cos (\theta _{ur})$. The angle of incident $\theta _{ur}$ can be found as follows. The halo mass $m_h$ satisfies
\begin{equation} 
\label{eq:10} 
m_{h} =\frac{4}{3} \pi r_{h}^{3} \Delta _{c} \bar{\rho } \Rightarrow  u_{cir} = \sqrt{\frac{Gm_{h} }{r_{h} }}=Hr_{h} \sqrt{\frac{\Delta _{c} }{2}}=3\pi u_r, 
\end{equation} 
where $r_h$ is the virial size of the halo, $\Delta _{c}=18\pi ^{2}$ is the critical halo density from a spherical collapsed model \citep{Gunn:1977-Massive-Galactic-Halos--1--For}, and $u_{cir}$ is the circular velocity at the boundary of halo. The Hubble parameter $H^2={8\pi G\bar{\rho }/3}$, where $\bar{\rho }\left(t\right)$ is the mean density of matter. Due to the stable clustering hypothesis for virialized haloes, there is no stream motion between particles in physical coordinates. In this sense, the peculiar motion cancels out the Hubble flow, and the hypothesis equivalently states that the peculiar radial velocity $u_{r}=Hr_{h}$ in Eq. \eqref{eq:10}. Therefore, the angle of incidence reads (on average)
\begin{equation} 
\label{ZEqnNum737533} 
\cos \left(\theta _{ur} \right) \approx \cot \left(\theta _{ur} \right)=\frac{u_{r} }{u_{cir} } = \frac{1}{3\pi }.         
\end{equation} 
Figure \ref{fig:8} represents a simple picture at the moment of a two-body merging, from which we hope to gain more insight.

From the above discussion, an alternative approach is presented here to estimate the constant rate of the energy cascade $\varepsilon_u$ that should give the same value as we obtained from the energy cascade in mass space (\ref{fig:S1-1-6}). The inverse cascade of mass and kinetic energy is facilitated by a series of halo merging with single mergers (Fig. \ref{fig:S2}). Upon each merging event, the mass is cascaded from a smaller scale $m_{h} $ to a larger scale $m_{h}+m$. Simultaneously, associated with the mass cascade, the kinetic energy is also transferred from the smaller to the larger mass scales that come from the change in the kinetic energy of that single merger at the instant of merging. Only relative motion along the radial direction is relevant for the energy cascade (Fig. \ref{fig:8}), which is due to the intra-halo interaction on small scales. Motion in the tangential direction does not contribute to the energy cascade. Therefore, for a single merger with a critical velocity $u_c$ and a critical acceleration $a_c$, the rate of the energy cascade $\varepsilon_{u}$ can be simply treated as the change in specific kinetic energy $u_r^2/2$ of that single merger, i.e.,  
\begin{equation} 
\label{ZEqnNum219659} 
\varepsilon _{u} \propto -\frac{1}{2}\frac{d(u_r^2)}{dt}=-a_{r} u_{r} =-a_c u_c \cos^2 \left(\theta _{ur} \right),  
\end{equation} 
where $\varepsilon _{u}<0$ for inverse cascade of kinetic energy from small to large scales, and $\cos(\theta_{ur})\approx \cot(\theta_{ur})$ for $\theta_{ur}$ close to $\pi/2$ in Eq. \eqref{ZEqnNum737533}. Here, the critical velocity and acceleration, or the scales of the velocity and acceleration fluctuations, are $u_c\equiv\lvert\boldsymbol{\mathrm{u}}\rvert$=std$(v_{hp}^i)$ and $a_c\equiv\lvert\boldsymbol{\mathrm{a}}\rvert$=std$(a_{hp}^i)$. Figures \ref{fig:5} and \ref{fig:6} present the redshift evolution of $u_c$ and $a_c$. This picture also reveals that the energy cascade in the dark matter flow is directly associated with the mass cascade. The mass and energy cascade are tightly coupled together \citep{Xu:2021-Inverse-mass-cascade-mass-function}. In contrast, the energy cascade in turbulence is facilitated by the deformation of the vortex structure, that is, by a vortex stretching mechanism \citep{Taylor:1938-Production-and-dissipation-of-}. Mass cascade does not exist in incompressible turbulence.

Combining Eqs. \eqref{ZEqnNum737533} and \eqref{ZEqnNum219659} together, the critical acceleration $a_c\left(a\right)$ can be related to the rate of the cascade $\varepsilon _{u}$ as
\begin{equation} 
\label{ZEqnNum138201} 
a_{c} \left(a\right) \approx -\Delta_c \frac{\varepsilon _{u} }{u_c} = -18\pi^2 \frac{\varepsilon _{u} }{u_c}\propto a^{-3/4} \propto t^{-1/2}.   \end{equation} 
With $u_{c} =496$km/s at $z=0$ from Fig. \ref{fig:6} and the constant rate of the energy cascade $\varepsilon_u=-3\times 10^{-7}m^2/s^3$ from Fig. \ref{fig:S1-1-6}, the critical scale $a_c$ for acceleration fluctuation reads
\begin{equation} 
\label{eq:15} 
\begin{split}
&a_{c}\left(z\right)=a_{c0}a^{-3/4}\propto a_{c0} t^{-1/2}, \quad \textrm{where} \\ 
&a_{c0}\equiv a_{c}\left(z=0\right)\approx 1.1\times 10^{-10} {m/s^{2}},
\end{split}
\end{equation} 
where $a_{c0}$ is the critical acceleration at $z=0$. Here, the velocity fluctuations and the acceleration fluctuations are related to each other through the rate of the energy cascade $\varepsilon_u$. Furthermore, the redshift dependence of the critical acceleration $a_c$ is predicted to be $a_{c} \left(a\right)\propto a^{{-3/4} } =\left(1+z\right)^{{3/4}}$ with a value of $a_{c}=1.1\times 10^{-10} m/s^{2}$ at $z=0$, which can be clearly confirmed by the N-body simulations in Fig. \ref{fig:5}. 

\begin{figure}
\includegraphics*[width=\columnwidth]{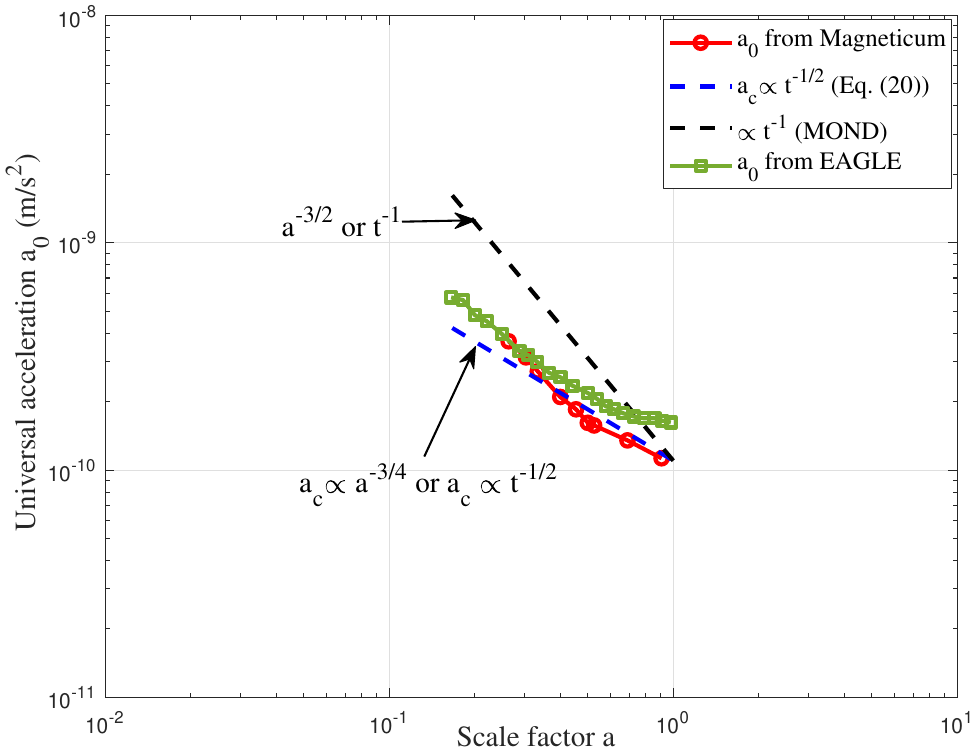}
\caption{The variation of the universal acceleration $a_0$ extracted from a large sample of galaxies in cosmological hydrodynamical Magneticum simulations (red circles) \citep{Mayer:2022-CDM-with-baryons-vs-MOND} and the EAGLE (green squares) \citep{Dai:2017-Can-the-Lambda-CDM-model-repro}. The predicted critical acceleration $a_c$ of dark matter varies with time as $a_c\propto a^{-3/4} \propto t^{-1/2}$ in Eq. \eqref{eq:15}. The critical acceleration $a_c$ matches the universal MOND acceleration $a_0$ extracted from Magneticum, which suggests the potential relations between two accelerations (Section \ref{sec:8}). The time evolution $a_0 \propto t^{-1}$ suggested by MOND was also plotted for comparison, which seems not to agree with the simulation. Our model predicts $a_0 \propto t^{-1/2}$ that matches better with the N-body hydrodynamic simulations.} 
\label{fig:S1-1-9}
\end{figure}

\section{Simulations and observations for redshift evolution of \texorpdfstring{$\MakeLowercase{a}_c$}{}}
\label{sec:8-2}
Return to the baryonic Tully-Fisher relation (BTFR in Eq. \eqref{eq:6-1}), the empirical acceleration $a_0\approx 10^{-10} m/s^2$ matches the value of the critical acceleration $a_c$ of dark matter in Eq. \eqref{eq:15}. Over the years, Modified Newtonian Dynamics (MOND) has been proposed to explain the empirical BTFR, where $a_0$ is a universal acceleration scale in MOND. In this section, we show that the critical acceleration $a_{c0}$ from the acceleration fluctuation matches the empirical acceleration $a_0$ in BTFR and MOND. More importantly, the predicted redshift dependence of $a_c\propto a^{-3/4}$ is consistent with the evolution of $a_0$ from hydrodynamic simulations, as shown in Fig. \ref{fig:S1-1-9}. 

For example, in Magneticum simulation \citep{Mayer:2022-CDM-with-baryons-vs-MOND}, the relation between the baryonic acceleration and the total acceleration, i.e., the rotational acceleration relationship (RAR), can be extracted from a large sample of galaxies. Universal acceleration $a_0$ can be identified from the RAR relation at different redshifts $z$. Figure \ref{fig:S1-1-9} presents the redshift variation of the universal acceleration $a_0$ from the Magneticum simulations (red circle). Our theory predicts the critical acceleration of dark matter $a_c\propto a^{-3/4}\propto t^{-1/2}$ (blue dashed line for Eq. \eqref{eq:15}) that matches the acceleration $a_0$ obtained from the hydrodynamic simulations. Furthermore, the redshift dependence of $a_0$ is obviously shallower than $\propto t^{-1}$ (black dashed line), which is proposed from the coincidence in MOND theory \citep{Milgrom:2001-MOND---A-pedagogical-review}, that is, $a_0=cH/(2\pi) \propto t^{-1}$, where $c$ is the speed of light. This conclusion can also be confirmed by the acceleration $a_0$ obtained from the EAGLE simulation results (green squares) \citep{Dai:2017-Can-the-Lambda-CDM-model-repro}, which is consistent with the scaling $a_c\propto a^{-3/4} \propto t^{-1/2}$. The larger $a_c$ at higher redshift means galaxies of fixed mass rotate faster at higher redshift with the scaling $v_f\propto (1+z)^{3/16}$ from Eq. \eqref{eq:6-1}. 

\rev{The Millennium Simulation with a large sample of $3\times 10^7$ galaxies of various morphologies also confirms this redshift evolution of the BTFR. In panel (c) of Fig. 14 in \cite{Obreschkow:2009-SIMULATION-OF-THE-COSMIC-EVOLUTION-OF-ATOMIC}, when comparing against the rotation at z=0, the galaxies of baryonic mass $M_b=4\times 10^6M_{\odot}$ rotate x1.4 faster at z=4.89 and x1.6 faster at z=10.07, consistent with the scaling of $v_f\propto (1+z)^{3/16}$. This also confirms a larger $a_0$ at a higher redshift, consistent with our predictions.} 

Limited observational evidence also exists for galaxies of fixed mass rotating faster at higher redshifts \citep{Sharma:2024-Tully-Fisher-relation-of-late-type-galaxies,Ubler:2017-The-Evolution-of-the-Tully,Straatman:2017-ZFIRE-The-Evolution-of-the-Stellar-Mass}. We provide a brief discussion here by generalizing the original BTFR in Eq. \eqref{eq:6-1} to include the redshift dependence, 
\begin{equation} 
\label{eq:6-2} 
\begin{split}
\log M_b = \alpha \log v_f -\gamma \log(1+z) + \beta_0, \\
\end{split}
\end{equation}
where $\alpha$ is a slope parameter for BTFR. Here, $\beta_0$ is an intercept parameter at $z=0$ that can be related to the critical acceleration at $z=0$ (Eq. \eqref{eq:6-3}. Redshift parameter $\gamma$ is introduced for the redshift dependence of the intercept of BTFR. For galaxies of fixed mass $M_b$, the larger $\gamma$ means the stronger redshift dependence and the larger rotation velocity $v_f$ at a high redshift. 

\begin{figure}
\includegraphics*[width=\columnwidth]{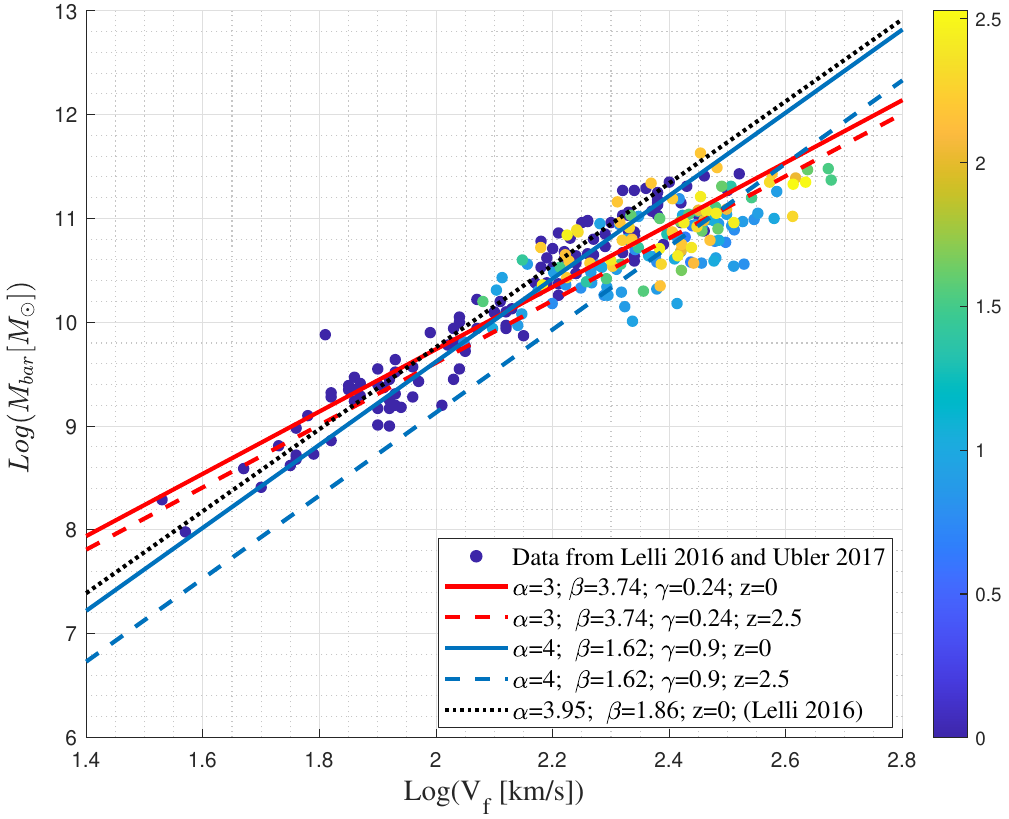}
\caption{The BTFR for our 253 samples of galaxies color-coded by the redshift of each galaxy. The first 118 galaxies are taken from \cite{Lelli:2016-THE-SMALL-SCATTER-OF-THE-BARYONIC} at $z=0$ (blue dots). The other 135 galaxies are taken from \cite{Ubler:2017-The-Evolution-of-the-Tully} with various redshift $0.6<z<2.5$. The dotted line plots the best fit of 118 local galaxies from \cite{Lelli:2016-THE-SMALL-SCATTER-OF-THE-BARYONIC}. The red solid and dashed lines plot the best free fit of all 253 galaxies at z=0 and 2.5, respectively, when including $\gamma$ for redshift dependence (Eq. \eqref{eq:6-2}), with best $\gamma=0.24$. The blue solid and dashed lines plot the best fit of 253 galaxies with a fixed slope $\alpha=4$, where the best $\gamma=0.9$. Relevant parameters ($\alpha$, $\beta_0$, and $\gamma$) can be found from black dashed lines in Fig. \ref{fig:S1-124}. Both cases confirm a positive $\gamma$ such that galaxies of fixed mass rotate faster at higher redshift. Larger slope $\alpha$ exhibits stronger redshift dependence.} 
\label{fig:S1-123}
\end{figure}

Specifically, for slope parameter $\alpha=4$ in Eq. \eqref{eq:6-1}, 
\begin{equation}
\label{eq:6-3} 
\alpha=4 \quad \beta_0=-\log\left(\frac{Ga_{c0}M_\odot}{10^{12}m^4/s^4}\right)=1.8.
\end{equation}
We assume a general power-law redshift dependence of the critical acceleration $a_0\propto (1+z)^\gamma$. Our theory predicts $\gamma=3/4$ if $a_0$ represents the acceleration fluctuations in dark matter, while MOND predicts either $\gamma=3/2$ if $a_0\propto cH$ (Fig. \ref{fig:S1-1-9}) or $\gamma=0$ if $a_0\propto cH_0$ is independent of $z$. We expect to estimate the value of $\gamma$ from local and high-redshift observations. 

\begin{figure}
\includegraphics*[width=\columnwidth]{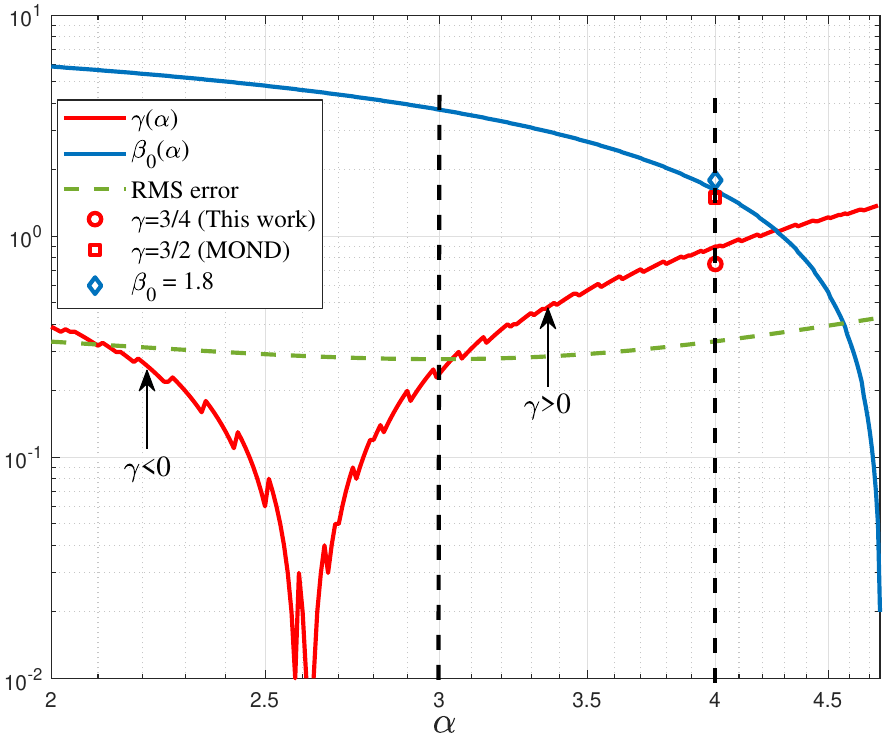}
\caption{The variation of the redshift parameter $\gamma$ and intercept parameter $\beta_0$ in Eq. \eqref{eq:6-2} fitting to all 253 galaxies for different fixed slope parameter $\alpha$. The RMS error of fitting is plotted as a dashed line. The BTFR with minimum RMS error for free fit among all different $\alpha$ is found at $\alpha=3$ and plotted as red lines in Fig. \ref{fig:S1-123}. The BTFR for fixed $\alpha=4$ is plotted as blue lines in Fig. \ref{fig:S1-123}. The redshift parameter $\gamma$ is positive between a range of $3<\alpha<4$, confirming a redshift-dependent BTFR. The larger $\alpha$, the larger $\gamma$, and the stronger redshift dependence of BTFR. At $\alpha=4$, the predicted $\beta_0$ is around 1.8 (diamond from Eq. \eqref{eq:6-3}). The predicted values of $\gamma=3/4$ from this work (red circle) and 3/2 from MOND (red square) are also shown. From these data, the critical acceleration $a_0$ is redshift dependent, with a smaller dependence than MOND prediction $a_0\propto (1+z)^{3/2}$, and in better agreement with simulations (Fig. \ref{fig:S1-1-9}) and our model of $a_0\propto (1+z)^{3/4}$.} 
\label{fig:S1-124}
\end{figure}

Figure \ref{fig:S1-123} presents the observed baryonic mass $M_b$ and rotation velocity $v_f$ for 118 local galaxies at $z=0$ taken from \cite{Lelli:2016-THE-SMALL-SCATTER-OF-THE-BARYONIC} and 135 high-redshift galaxies with various redshift $0.6<z<2.5$ taken from \cite{Ubler:2017-The-Evolution-of-the-Tully} (symbols colored by redshift). The standard approach to obtain the BTFR is the two-parameter fit of slope $\alpha$ and intercept $\beta_0$ \citep{Lelli:2016-THE-SMALL-SCATTER-OF-THE-BARYONIC,Sharma:2024-Tully-Fisher-relation-of-late-type-galaxies,Ubler:2017-The-Evolution-of-the-Tully}. In contrast, we perform the least-squares fit of three parameters $\alpha$, $\beta_0$, and $\gamma$ in Eq. \eqref{eq:6-2} for all 253 galaxies to evaluate the redshift dependence. For every fixed slope $\alpha$, Fig. \ref{fig:S1-124} presents the best fitting of $\gamma$ and $\beta_0$ in Eq. \eqref{eq:6-2}. The corresponding RMS error of fitting is also presented. The error slowly increases toward $\alpha=4$ with the minimum error obtained at $\alpha=3$ (dashed black line). The corresponding BTFR for $\alpha=3$ and 4 were also presented in Fig. \ref{fig:S1-123} (red and blue lines). Both figures confirm a redshift-dependent BTFR such that galaxies of fixed mass rotate faster at higher redshifts. Data suggests a positive $\gamma$ increasing with $\alpha$. More importantly, for $\alpha=4$, observation data also suggests $\gamma$ is smaller than the MOND prediction ($\gamma=3/2$) and is in better agreement with our prediction $\gamma=3/4$, which supports the origin of critical acceleration $a_0$ from acceleration fluctuations. Future study requires more high quality high-redshift data.

The time evolution of $a_c\propto (1+z)^{3/4}\propto t^{-1/2}$ in Eq. \eqref{eq:15} offers a powerful test to distinguish between MOND and $\Lambda$CDM with a redshift dependent $a_c$. The coincidence between the critical acceleration of dark matter $a_{c}$ and the universal acceleration $a_0$ in BTFR and MOND suggests a potential connection between two accelerations. In the next Section, we will briefly discuss this connection. 


\section{The origin of universal acceleration \texorpdfstring{$\MakeLowercase{a}_0$}{}}
\label{sec:8}
 Similarly to the critical velocity scale $u_c$ due to the fluctuation of the velocity, there exists a critical acceleration $a_c$ due to the fluctuation of the acceleration (Figs. \ref{fig:4} and \ref{fig:5}). Note that the critical acceleration $a_c$ at \textit{z}=0 matches the empirical acceleration $a_0$ in BTFR and also the universal MOND acceleration. This suggests that $a_{0} $ could be an intrinsic property of $\Lambda$CDM cosmology due to the acceleration fluctuations in dark matter. The value and origin of $a_{0}$ in BTFR is still empirical and phenomenological without a good theory. In this section, we briefly discuss the connection between empirical acceleration $a_0$ and critical acceleration $a_c$ in dark matter. 

First, the modified Newtonian Dynamics (MOND) is a popular empirical model that reproduces the same astronomical observations without invoking the dark matter hypothesis \citep{Milgrom:1983-A-Modification-of-the-Newtonia}. The basic idea of MOND is to introduce a universal acceleration $a_0$. Standard Newtonian mechanics $F=ma$ is recovered when baryon acceleration $a\gg a_{0}$. For the "deep-MOND" regime with $a\ll a_{0}$, Newtonian mechanics should be modified to $F=m{a^{2}/a_{0}}$, that is, the external force $F\propto a^{2}$. 
In the "deep-MOND" regime, for a galaxy with a total baryonic mass $M_b$, the dynamics for a given baryon particle far from the center of the galaxy is given by (with acceleration $a=v_f^2/r$, where $r$ is the distance):
\begin{equation}
\label{ZEqnNum417694} 
\frac{F}{m}=\frac{GM_{b} }{r^{2} } =\frac{({v_{f}^{2}/r} )^{2} }{a_{0} }  \Rightarrow  v_{f}^{4} =GM_{b} a_{0} ,       
\end{equation} 
which is the BTFR with a rotation velocity $v_f$ in Eq. \eqref{eq:6-1}. 

The universal MOND acceleration $a_0$ (or the empirical BTFR acceleration) might originate from the critical acceleration $a_c$ due to fluctuations in dark matter acceleration. Let us consider baryons dispersing and mixing with a fluctuating dark matter fluid. The self-gravitating collisionless dark matter fluid has an intrinsic critical velocity scale $u_{c}$ and a critical acceleration scale $a_{c}$ due to respective velocity and acceleration fluctuations (Sections \ref{sec:5} and \ref{sec:4}). The rate of the energy cascade can be related to the velocity and acceleration fluctuations as $\varepsilon_u \propto -a_c u_c$ (Eq. \eqref{ZEqnNum219659}) (Section \ref{sec:6}). For the entire halo, the total mass of dark matter is dominant over the mass of baryons. However, in the core region of haloes (bulge), the baryonic mass (star plus gas) can be dominant over the mass of dark matter. 

Consider baryonic particles on the outskirts of the halo with mass $m_{b}$, velocity $u_{b}$, and acceleration $a_{b}={du_{b}/dt}$, moving under an external driving force $F_{b}$. The baryons interact with dark matter through gravity only so that the baryons in the halo outskirts are in equilibrium with dark matter and share the same rate of energy cascade such that $\varepsilon_u \propto -a_b u_b$ (Eq. \eqref{ZEqnNum605626}). 
The acceleration of baryons on the outskirts can be extremely small and less than the critical acceleration $a_c$ of the fluctuation, that is, $a_b < a_c$. 
In this scenario, the effect of fluctuations is dominant over the effect of external force $F_b$. The acceleration of the baryon particle $a_b$ and the velocity $u_b$ should give the same rate of the energy cascade:
\begin{equation} 
\label{ZEqnNum605626} 
\begin{split}
-\varepsilon_{u} \propto a_c u_c = a_{b} u_{b}.     
\end{split}
\end{equation} 

Next, the external force $F_{b}$ can be calculated from the time derivative of the specific kinetic energy $\varepsilon_K$. This means the change in the particle kinetic energy equals the power from the external force. Therefore, we write 
\begin{equation} 
\label{ZEqnNum768079} 
\begin{split}
&F_b u_b = m_b \frac{d\varepsilon_{K}(u_b)}{dt}=m_b \frac{d\varepsilon_{K}(u_b)}{d u_b} \frac{d u_b}{d t}=m_bu_ca_b,  \\
\end{split}
\end{equation} 
where $\varepsilon_K \propto u_c u_b$ from Eq. \eqref{ZEqnNum201633}), i.e., the kinetic energy of baryon particles shares the same form as dark matter particles. Here, the left-hand side is the input of power (the work done by $F_b$ per unit of time) from the external force $F_b$. The right-hand side shows the change in the kinetic energy of baryons. The work done by $F_b$ increases the kinetic energy $\varepsilon_K$ of the baryon. In self-gravitating collisionless dark matter flow, the effective kinetic energy $\varepsilon_K$ can be obtained from the maximum entropy distributions \citep{Xu:2023-Maximum-entropy-distributions}. In Eq. \eqref{ZEqnNum201633}, the effective kinetic energy follows $\varepsilon_{K} \left(v\right)\propto v^{2} $ for particles in the core region with low speed or high acceleration ($v\ll v_0$), that is, the standard Newtonian behavior. However, $\varepsilon _{K} \left(v\right)\propto v_{0} v$ for high-speed or low-acceleration particles ($v\gg v_0$) in the outskirts of haloes is a unique "non-Newtonian" feature due to the long-range nature of gravity. Here, the velocity scale $v_0$ plays the role of the critical velocity $u_c$. 

Finally, for baryon particles with $a_b \ll a_c $ or $u_b \gg u_c$ (in the outskirts region of the halo), the specific kinetic energy $\varepsilon _{K}\propto u_c u_b$ due to the long-range gravity and inter-halo interactions. Substituting this into Eq. \eqref{ZEqnNum768079}, the external force can be calculated from Eqs. \eqref{ZEqnNum605626} and \eqref{ZEqnNum768079} as
\begin{equation} 
\label{ZEqnNum666612} 
F_b = m_b a_b \frac{u_c}{u_b} = m_b \frac{a_b^2}{a_c}. 
\end{equation} 
This result is in agreement with the force in the "deep-MOND" regime. Compared Eq. \eqref{ZEqnNum666612} against Eq. \eqref{ZEqnNum417694}, we found that the critical acceleration $a_c$ due to acceleration fluctuations in dark matter plays the role of universal acceleration $a_0$ in BTFR or MOND. Furthermore, the time evolution of $a_c\propto t^{-1/2}$ in Eq. \eqref{eq:15} suggests the same time evolution of $a_0\propto t^{-1/2}$. This theory offers a powerful test to distinguish between MOND and $\Lambda$CDM from observations of the redshift variation of $a_0$. 

On the other hand, for particles with high acceleration in the halo core region, that is, $a_b\gg a_c$, the effect of external force $F_b$ is dominant over fluctuations. The intra-halo interactions are dominant over inter-halo interactions. The standard Newton law $F_b=m_b a_b$ can be recovered by inserting $\varepsilon_K \propto u_b^2$ from Eq. \eqref{ZEqnNum201633} into Eq. \eqref{ZEqnNum768079}. 

In short, two key ingredients are necessary for this simple interpretation: i) the constant rate of the energy cascade for both dark matter and baryons (Eq. \eqref{ZEqnNum605626}), and ii) the effective kinetic energy proportional to the speed at a small acceleration (Eq. \eqref{ZEqnNum201633}). With these two ingredients, we may recover the MOND. In conventional wisdom, MOND is a competing empirical theory that falsifies dark matter. In this work, we propose that MOND is an intrinsic feature of and consistent with $\Lambda$CDM. Instead of falsifying dark matter, MOND is an effective theory for the dynamics of baryons in equilibrium with self-gravitating collisionless dark matter. 

\section{Relevant physical quantities on small and large scales}
\label{sec:7}
Figure \ref{fig:S2} discusses the propagation range between the smallest mass scale $m_X$ and the characteristic halo mass $m_h^*$. The smallest mass scale depends on the particle mass and free streaming. The smallest structure that can be formed by dark matter particles of any mass has a mass around $m_X=10^{12}$GeV \citep{Xu:2022-Postulating-dark-matter-partic}. The propagation range involves a constant rate of cascade $\varepsilon_u$, which is independent of the mass scale. The value of $\varepsilon_u$ is determined in Section \ref{sec:6-1} (Fig. \ref{fig:S1-1-6}). The velocity and acceleration fluctuations introduce two additional quantities, i.e. the velocity scale $u_c$ and acceleration scale $a_c$, in the propagation range. In this section, similarly to the application of dimensional analysis in fluid dynamics and turbulence, we will apply dimensional analysis to the flow of collisionless dark matter to determine relevant physical quantities at the small end $m_X$ and the large end $m_h^*$ of the propagation range. 

First, let us focus on the large end in Fig. \ref{fig:S2}. This is the largest halo scale at $z=0$ (denoted by the subscript $A$). On that scale, the expansion of the universe is important. The Hubble constant $H_0$ and the fluctuation scales ($u_c$ and $a_c$) should play a role on this scale. The corresponding one-dimensional velocity scale is $u_0=u_c/\sqrt{3}$. The dominant physical constants on this scale are $G$, $\varepsilon_{u}$, the critical velocity scale $u_0\approx 286$km/s from Fig. \ref{fig:6} (or the acceleration fluctuation scale $a_c$), and the scale factor $a$ (or the Hubble parameter $H$). Any physical quantity $Q$ on this scale can be expressed as $Q \propto G^x\varepsilon_{u}^yu_0^za^p$, where the exponents (x, y, z) can be determined from the dimensional analysis. 

Table \ref{tab:2} lists the relevant physical quantities on the largest scale $A$ (at $z=0$) calculated with values for $\varepsilon_u$, $G$ and $u_0$:
\begin{equation} 
\label{ZEqnNum7680799} 
\begin{split}
&\varepsilon _{u} =-3\times 10^{-7} {m^{2} /s^{3} },\\
&G=6.67\times 10^{-11} {m^{3} /(kg\cdot s^{2})}, \\ 
&u_0 = 286km/s.\\ 
\end{split}
\end{equation} 
The length scale $l_A\propto u_0^3/\varepsilon_u$=2.5Mpc is about the size of the largest halo with a characteristic halo mass $m_A \approx m_h^*\approx 10^{13}M_{\odot}$. The time scale $t_A$ should be the time (approximately the age of the universe $t_0$) required to form the largest halo. The density scale $\rho_A\propto \varepsilon_u^2/Gu_0^4$ is the average mass density of the entire halo. If we write the mean dark matter density as ($\Omega_m$ is the mass fraction of dark matter),
\begin{equation}
\label{eq:23-13}
\begin{split}
\bar{\rho}=\frac{3H_0^2}{8\pi G} \Omega_m \approx 2\times 10^{-27} kg/m^3, 
\end{split}
\end{equation}
the density contrast $ \Delta_c = \rho_A/\bar\rho$ can be obtained as 
\begin{equation}
\label{eq:23-14}
\begin{split}
\Delta_c = \frac{\rho_A}{\bar{\rho}} =\frac{8\pi}{3\Omega_m} \left(\frac{\varepsilon_u}{H_0u_0^2}\right)^2 \approx 100.
\end{split}
\end{equation}
This density contrast is of the same order as the value $\Delta_c=18\pi^2$ obtained independently from the spherical collapse model \citep{Gunn:1972-Infall-of-Matter-into-Clusters}. 
The acceleration scale $a_A$ has already been discussed in detail (Eq. \eqref{ZEqnNum138201}), which is the critical acceleration in fluctuation.

The energy density scale $\rho_{eA}$ is comparable to the energy density of CMB photons (cosmic microwave background). The CMB photons interact with baryons via Thomson scattering, whereas dark matter interacts with baryons via gravity. With all three "phases" (baryons, radiation, and dark matter) in equilibrium on the largest halo scale, the energy density or pressure of the haloes on that scale $\rho_{eA}$ should be comparable to that of the CMB density $\rho_{CMB} \approx 4\times 10^{-14}J/m^3$, that is, the halo pressure on that scale is in balance with the background radiation. Finally, the "effective" kinematic viscosity $\nu_A$ is relevant to the adhesion model in large-scale structure formation \citep{Gurbatov:1989-The-Large-Scale-Structure-of-t,Xu:2024-On-the-statistical-theory-of-self-gravitating}.

The small mass end of the propagation range (or mass scale $m_X$) was discussed in a separate paper \citep{Xu:2022-Postulating-dark-matter-partic}. For cold dark matter that is fully collisionless, the scale-independent cascade rate $\varepsilon_u$ (Fig. \ref{fig:S1-1-6}) can be extended down to the smallest mass scale $m_X$, where quantum effects become important if there are no other known interactions involved except gravity. This is the field of non-relativistic quantum gravity or Newtonian quantum gravity. Assuming that gravity is the only interaction between dark matter particles (traditionally denoted by \textbf{\textit{X}}), the dominant constants on the smallest scale (the $X$ scale) are the (reduced) Planck constant $\hbar$, the gravitational constant $G$, and the rate of the energy cascade $\varepsilon_{u}$. Similarly to the large end, any physical quantity $Q$ on this scale can be expressed as $Q \propto G^x\varepsilon_{u}^y\hbar^z$, where exponents can be determined from the dimensional analysis. With values listed in Eq. \eqref{ZEqnNum7680799} and $\hbar=1.05\times 10^{-34}kg\cdot m^2/s$, all relevant quantities on the mass scale $m_X$ can easily be found and listed in Table \ref{tab:2}. Details can be found in relevant references \citep{Xu:2023-Universal-scaling-laws-and-density-slope, Xu:2022-Postulating-dark-matter-partic}.

\begin{table}
\footnotesize
\centering
    \begin{tabular}{ccc}
    \hline
    \makecell{Quantity} &  \makecell{The smallest scale $X$} & \makecell{The largest scale $A$ \\ of halo at $z=0$}   \\
    \hline
    \makecell{Length}   &\makecell{$l_{X} =\left(-{G\hbar }/{\varepsilon_{u}}\right)^{{1}/{3}}$
                    \\=$10^{-13} \textrm{m}$}   & \makecell{$l_A =-u_0^3/\varepsilon_{u}$ 
                    \\=$2.5 \textrm{Mpc}$}\\
    \hline
    \makecell{Time}     &\makecell{$t_{X} =\left(-{G^{2}\hbar^{2}}/{\varepsilon _{u}^{5}}\right)^{{1}/{9}}$                             \\=$10^{-7} \textrm{s}$}       & \makecell{$t_A =\left(-u_0^2\varepsilon_{u}\right)$ 
                    \\=$8.6\times 10^9 \textrm{year}$}\\
    \hline
    \makecell{Mass}     &\makecell{$m_{X} =\left(-{\varepsilon_{u}\hbar^{5}}/{G^{4}}\right)^{{1}/{9}}$
                    \\=$10^{12} \textrm{GeV}$ \\}    & \makecell{$m_A =-u_0^5/(G\varepsilon_{u})$ 
                    \\=$4.8\times 10^{13} {M_{\odot}}$\\}\\
    \hline
    \makecell{Velocity} &\makecell{$v_{X} =\left({\varepsilon_{u}^{2}\hbar G}/{4}\right)^{{1}/{9}}$
                    \\=$10^{-7} \textrm{m/s}$}  & \makecell{$v_A =u_0$ 
                    \\= $286 \textrm{km/s}$}\\
    \hline
    \makecell{Accele\\-ration} &\makecell{$a_{X} =\left(-{\varepsilon_{u}^{7}}/{(\hbar G)}\right)^{{1}/{9}}$
                    \\$=1.11 {m/s^2}$}             & \makecell{$a_A =a_c=-\Delta_c {\varepsilon_{u}}/u_c$
                    \\=$1.1\times 10^{-10} {m/s^2}$ \\ (see Eq. \eqref{ZEqnNum138201})}\\
    \hline
    \makecell{Energy} &\makecell{$E_{X} =\left(-{\hbar^7\varepsilon_{u}^5}/G^2\right)^{{1}/{9}}$
                    \\=$10^{-9} \textrm{eV}$}     & \makecell{$E_A =-u_0^7/({G\varepsilon_{u}})$
                    \\=$7.8\times 10^{54} \textrm{J}$}\\
    \hline
    \makecell{Density} &\makecell{$\rho_{X} =\left({\varepsilon_{u}^{10}}/{(\hbar^4 G^{13})} \right)^{{1}/{9}}$                         \\=$10^{22} {kg/m^3}$}  & \makecell{$\rho_A ={\varepsilon_{u}^{2}}/{(G                               u_0^{4})}$\\=$2\times 10^{-25} {kg/m^3}$}    \\
    \hline
    \makecell{Energy \\density\\(pressure)}  &\makecell{$\rho_{eX} =\left(\varepsilon_{u}^{14}\hbar^{-2}G^{-11}\right)^{{1}/{9}}$                          \\=$10^{10} {J/m^3}$}  & \makecell{$\rho_{eA} = \varepsilon_{u}^{2}/{(Gu_0^2)}$                          \\=$1.6\times 10^{-14} {J/m^3}$} \\
    \hline
    \makecell{Diffusivity\\(kinematic \\ viscosity)} &\makecell{$\nu_{X} =\left(-{\hbar^4 G^{4}}\varepsilon_{u}^{-1}\right)^{{1}/{9}}$
                    \\=$10^{-19} {m^2/s}$}  & \makecell{$\nu_A =-u_0^4/\varepsilon_{u}$
                    \\=$2.2\times 10^{28} {m^2/s}$}\\
    \hline
    \end{tabular}
    \caption{Relevant physical quantities on large and small ends of the propagation range in Fig. \ref{fig:S2}.}
    \label{tab:2}
\end{table}

\section{Dark energy from acceleration fluctuations}
\label{sec:7-2}
With the critical acceleration scale $a_{c0}\approx 10^{-10}m/s^2$, an energy density can be naturally related to the acceleration fluctuation as ${a_{c0}^{2}/G}\approx 10^{-10}$ J/m$^3$, which is on the same order as the density of dark energy $\rho_{DE0}$ at $z=0$. In this section, we briefly discuss a possible model for dark energy as an energy density originating from the acceleration fluctuations of dark matter and compare this model with existing models. Discussions are only briefly presented here in the hope of leading to other suggestions and ideas from the community. Future studies require rigorous thermodynamic treatment for systems involving long-range interactions. 

Dark energy, a pervasive form of energy, was proposed to account for the accelerated expansion of the universe. The nature of dark energy has remained elusive for more than a quarter century. A key question of dark energy is whether it is a cosmological constant $\Lambda$, or it dynamically evolves, or it arises from the modifications of General Relativity. 

For dynamical models of dark energy, the first example is a flat $w$CDM model, where the evolution of the dark energy density $\rho_{DE}$ involves an equation of state parameter $w$ such that $\rho_{DE}\propto a^{-3(1+w)}$. Parameter $w$=-1 recovers the standard $\Lambda$CDM model with a constant dark energy density. The second example involves a two-parameter parametrization of the equation of state parameter $w(a)=w_0+w_a(1-a)$ ($w_0w_a$CDM model). This model leads to an evolving dark energy density of \citep{Tripathi:2017-Dark-energy-equation-of-state-parameter},
\begin{equation}
\label{eq:29}
\begin{split}
\rho_{DE} = \rho_{DE0} a^{-3(1+w_0+w_a)}e^{3w_a(a-1)},
\end{split}
\end{equation}
The recent results on baryon acoustic oscillations from DESI (Dark Energy Spectroscopic Instrument) provide constraints on two parameters $w_0$ and $w_a$ when combined with cosmic microwave background and supernova data \citep{desicollaboration:2024-Cosmological-Constraints}. These results prefer $w_0>-1$ and $w_a<0$ and indicate potential deviations from the standard $\Lambda$CDM model with a quickly weakening dark energy at low redshift due to the exponential term in Eq. \eqref{eq:29}. 

We begin with an analogy of the pressure in the kinetic theory of gases. The Maxwell-Boltzmann distribution of gas molecule velocity can be obtained from the maximum entropy principle (Eq. \eqref{ZEqnNum8641322}). There exist velocity fluctuations with a typical velocity scale $v_{gas}$. Kinetic pressure (or energy density) $P_{gas}$, a macroscopic state variable, is related to the velocity fluctuations (or temperature $k_BT\propto v_{gas}^2$),
\begin{equation}
\label{eq:23-1-1}
P_{gas} \propto \rho_{gas} v_{gas}^2, 
\end{equation}
where $\rho_{gas}$ is the density of gas. This suggests a fluctuation (an entropic or random motion) nature of the kinetic pressure $P_{gas}$. Thermal fluctuations tend to bring a system to its macroscopic state of maximum entropy with a finite pressure $P_{gas}$. The larger the velocity fluctuations $v_{gas}^2$, the higher the temperature $T$, the greater the entropy contained in the velocity distribution, and the higher kinetic pressure $P_{gas}$. 

For the self-gravitating flow of collisionless dark matter, acceleration fluctuations also exist with a typical scale $a_{c0}\equiv a_{c} \left(z=0\right)$, in addition to the velocity fluctuations (Sections \ref{sec:5} and \ref{sec:6}). Similarly to the kinetic pressure in Eq. \eqref{eq:23-1-1}, we postulate the existence of an additional energy density that is due to the acceleration fluctuations of all dark matter. Therefore, in an analog to Eq. \eqref{eq:23-1-1} for gas pressure, an acceleration-fluctuation-induced energy density $\rho_a$ at $z=0$ reads (using Eq. \eqref{ZEqnNum138201})
\begin{equation}
\label{eq:23-1}
\begin{split}
\rho_{a} = A_0 \frac{a_{c0}^2}{G} \approx 2A_0\times 10^{-10}{J}/{m^3},
\end{split}
\end{equation}
which coincidentally is on the same order of dark energy density $\rho_{DE0}=5\times 10^{-10}J/m^3$ at $z=0$. 
Here, the dimensionless numerical constant $A_0$ in Eq. \eqref{eq:23-1} is of unity. Unlike the kinetic pressure in Eq. \eqref{eq:23-1-1}, there is no density involved in Eq. \eqref{eq:23-1} such that this energy density due to acceleration fluctuation is independent of the local matter density. On large scales, it should be relatively homogeneous in space. 

Since the acceleration-fluctuation-induced energy density $\rho_a$, if exists, happens to be on the same order as the dark energy density at $z=0$ (Eq. \eqref{eq:23-1}), we postulate the dark energy with a possible origin from the acceleration fluctuation $\left\langle a_{p}^2 \right \rangle$ (Fig. \ref{fig:5}) such that the dark energy density reads
\begin{equation}
\label{eq:24}
\begin{split}
\rho_{DE} = A_0 \frac{\left\langle a_{p}^2 \right \rangle}{G} a^{-\nu_0},
\end{split}
\end{equation}
where $\left\langle a_{p}^2 \right \rangle$ is the acceleration fluctuations of all dark matter particles. Here, $\nu_0$ is an exponent of scale factor $a$ that can be constrained from observations. The evolution of the standard deviation of particle acceleration std($a_p$)=$\sqrt{\left\langle a_{p}^2 \right \rangle}$ is presented in Fig. \ref{fig:5} (magenta color).

Since all dark matter particles can be divided into the halo particles with a total mass of $M_h$ and out-of-halo particles, the fluctuations of all particles can be written as a weighted average involving a mass fraction $M_h/M$ (Here, $M$ is the total mass of all dark matter):
\begin{equation}
\label{eq:25}
\begin{split}
\left\langle a_{p}^2 \right \rangle = \left\langle a_{hp}^2 \right \rangle \frac{M_h}{M}+\left\langle a_{op}^2 \right \rangle \left(1-\frac{M_h}{M}\right) \approx \left\langle a_{hp}^2 \right \rangle \frac{M_h}{M}.
\end{split}
\end{equation}
We consider two phases for dynamical dark energy coupled to the halo structure formation and evolution in Eqs. \eqref{eq:24} and \eqref{eq:25}. At a very high redshift when the total mass in haloes is negligible, $M_h/M\approx 0$ such that $\left\langle a_{p}^2 \right \rangle \approx \left\langle a_{op}^2 \right \rangle \propto a^{-1}$ (Fig. \ref{fig:5}). In that situation, an exponent $\nu_0=-1$ is required for a constant dark energy density in Eq. \eqref{eq:24}. 

With the hierarchical formation of halo structures, the mass fraction $M_h/M\propto a^{1/2}$ increases with time and approaches one \citep{Xu:2023-Dark-matter-halo-mass-functions-and}.  Since the acceleration fluctuations of out-of-halo particles are much smaller than that of halo particles ($\left\langle a_{op}^2 \right \rangle\ll \left\langle a_{hp}^2 \right \rangle$ in Fig. \ref{fig:5}), we can make the approximation in Eq. \eqref{eq:25} where acceleration fluctuations of halo particles become dominant over out-of-halo particles. With $\left \langle a_{hp}^2 \right \rangle \propto a^{-3/2}$ and $M_h\propto a^{1/2}$, an exponent $\nu_0=-1$ is also required for a constant dark energy density in this phase.

Next, the acceleration fluctuations of halo particles read
\begin{equation}
\label{eq:26}
\begin{split}
\left\langle a_{hp}^2 \right \rangle=a_c^2\approx a_{c0}^2a^{-3/2},
\end{split}
\end{equation}
where $a_{c0}=a_c(z=0)\approx 10^{-10}m/s^2$ is the critical acceleration. The total mass of all haloes can be shown to follow a power-law scaling of $M_h\propto a^{1/2}$ in the matter era and reach a plateau in the dark energy era \citep{Xu:2022-Postulating-dark-matter-partic,Xu:2023-Dark-matter-halo-mass-functions-and,Xu:2021-Inverse-mass-cascade-mass-function}. Therefore, the mass fraction of all haloes in all dark matter can be approximately modeled as
\begin{equation}
\label{eq:27}
\begin{split}
\frac{M_h}{M} = \alpha_h a^{1/2} \left(1-\nu_a{a}\right)^{-1/2},
\end{split}
\end{equation}
where $\alpha_h\approx 0.6$ reflects that around 60\% of all dark matter are in haloes at $z=0$ from N-body simulations. At small $a$, the scaling $M_h\propto a^{1/2}$ is recovered, while at large $a$, the mass fraction $M_h/M$ levels off with a maximum mass fraction $\alpha_h(-\nu_a)^{-1/2}$ that is determined by the second parameter $\nu_a$. 

\begin{figure}
\includegraphics*[width=\columnwidth]{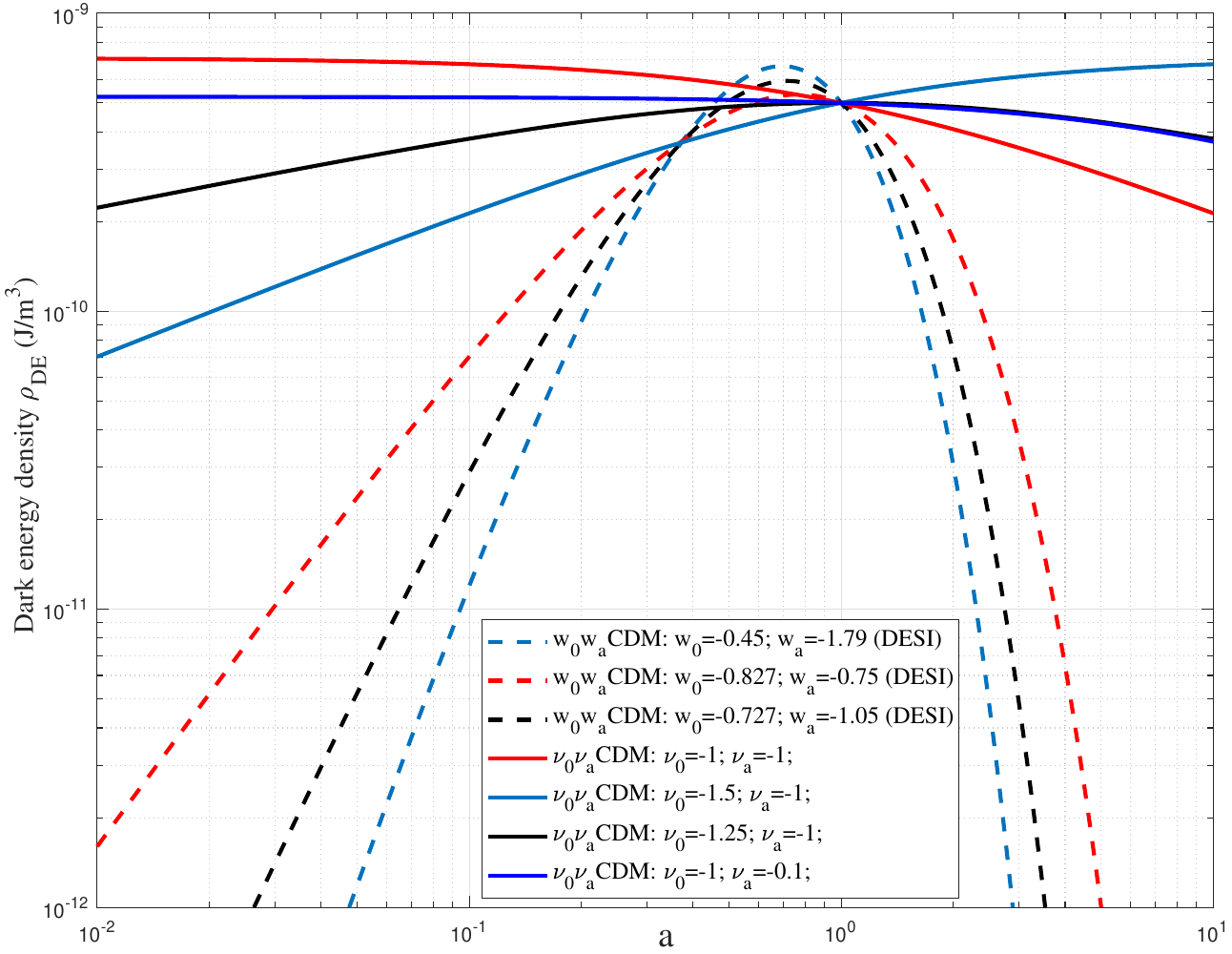}
\caption{The evolution of dark energy density from the $\nu_0\nu_a$CDM model (Eq. \eqref{eq:28}) and from the $w_0w_a$CDM model (Eq. \eqref{eq:29}). The $\nu_0\nu_a$CDM model proposes the evolution of dark energy is tightly coupled to the evolution of dark matter haloes. The model postulates an entropic origin of dark energy from acceleration fluctuations of dark matter, which naturally gives rise to the dark energy density $\rho_{DE0}\approx a_{c0}^2/G=10^{-10}$J/m$^3$. This model avoids the "cosmological constant problem" since dark energy is different from vacuum energy, the intrinsic energy of empty space. The model supports a dynamic evolution of dark energy with an almost constant dark energy density in the matter era, followed by a weakening dark energy when halo growth slows down. However, the $\nu_0\nu_a$CDM model predicts a much slower power-law weakening compared to the exponential decrease in the $w_0w_a$CDM model.}
\label{fig:8-3}
\end{figure}

Putting all equations together, the dark energy density from acceleration fluctuations of the dark matter finally reads
\begin{equation}
\label{eq:28}
\begin{split}
&\rho_{DE} = A_0 \frac{a_{c0}^2}{G} \alpha_h a^{-\nu_0-1}\left(1-\nu_a{a}\right)^{-1/2}, \\
&\textrm{or}\\
&\rho_{DE} = \rho_{DE0} a^{-\nu_0-1} \left(\frac{1-\nu_a}{1-\nu_a a}\right)^{1/2} \\
&\textrm{where}\quad \rho_{DE0}=A_0 \frac{a_{c0}^2}{G}\alpha_h\left(1-\nu_a\right)^{-1/2}.
\end{split}
\end{equation}
This is a double power-law evolution of dark energy density, i.e., $\rho_{DE}\propto a^{-\nu_0-1}$ in the matter era. With the total mass of all haloes $M_h$ leveling off with time, the dark energy density decreases as $\rho_{DE}\propto a^{-\nu_0-3/2}$ in the dark energy era. The standard $\Lambda$CDM model with a constant dark energy density is recovered with $\nu_0=-1$ and $\nu_a=0$. With $\nu_0=-1$, the model of Eq. \eqref{eq:28} suggests a constant dark energy density phase in the matter era followed by a slowly decreasing phase of dark energy density as $\rho_{DE}\propto a^{-1/2}$. Here, $\nu_a$ is a shape parameter that determines the transition between two phases. 

The proposed dark energy model involves two parameters $\nu_0$ and $\nu_a$ (i.e., the $\nu_0\nu_a$CDM model). Figure \ref{fig:8-3} presents the evolution of the $\nu_0\nu_a$CDM model. When compared against the $w_0w_a$CDM model with constraints from DESI \citep{desicollaboration:2024-Cosmological-Constraints}, the $\nu_0\nu_a$CDM model has three distinct features: i) the evolution of dark energy is tightly coupled to the evolution of dark matter haloes; ii) model postulates that dark energy originates from the acceleration fluctuations such that the "cosmological constant problem" can be avoided; iii) model suggests a constant dark energy density in the matter era and a weakening evolution in the dark energy era when structure formation slows down. However, the $\nu_0\nu_a$CDM model predicts a much slower power-law weakening compared to the exponential decrease in the $w_0w_a$CDM model.


\section{Conclusions}
\label{sec:9}
This paper focuses on the velocity and acceleration fluctuations in self-gravitating collisionless dark matter flow (SG-CFD) due to the long-range nature of gravity. Long-range interaction requires the formation of different sizes of haloes to maximize the entropy of the system \citep{Xu:2023-Maximum-entropy-distributions}. These haloes facilitate an inverse mass and energy cascade from small to large scales with a constant rate of the energy cascade $\varepsilon _{u}\approx -10^{-7}m^2/s^3$ \citep{Xu:2021-Inverse-mass-cascade-mass-function}. In addition to velocity fluctuations involving a critical velocity scale $u_c$, the long-range nature of gravity leads to fluctuations in acceleration with a critical scale $a_c$. The fluctuations in velocity and acceleration satisfy the relation $\varepsilon _{u} =-a_c u_c/(18\pi^2)$, which gives $a_c \approx 10^{-10} {m/s^{2}}$ for $u_c \approx 300$km/s at $z=0$. The critical acceleration scale $a_c$ matches the empirical acceleration $a_0$ in the baryonic Tully-Fisher relation (BTFR) and the universal acceleration in MOND. In conventional wisdom, MOND is a competing empirical theory that potentially falsifies the dark matter hypothesis. Instead of falsifying dark matter, we suggest that MOND theory might be consistent with $\Lambda$CDM cosmology due to the velocity and acceleration fluctuations and the energy cascade in cold and collisionless dark matter. The "deep MOND" behavior can be recovered on the basis of this idea. More importantly, the predicted redshift dependence of $a_c\propto (1+z)^{3/4}$ is in good agreement with hydrodynamic simulations and observations. This redshift-dependent acceleration $a_c$ offers a powerful test to distinguish MOND and $\Lambda$CDM. In analogy to the kinetic pressure of gases that are produced by and proportional to velocity fluctuations, we postulate a similar "pressure" or "energy density" $\rho_a$ that is induced by the acceleration fluctuations of dark matter. This energy density is proportional to acceleration fluctuations ($\rho_a\propto a_c^2/G$) and is of the order of $10^{-10}$ J/m$^3$, or on the same order as the density of dark energy. Therefore, we postulate an entropic origin of the dark energy from the acceleration fluctuations, where the evolution of dark energy is coupled to the structure formation and evolution. A dynamical dark energy model ($\nu_0\nu_a$CDM model in Eq. \eqref{eq:28}) was proposed with a constant dark energy density at high redshift followed by a slow weakening phase. 

\section*{Acknowledgments}
This research was supported by Laboratory Directed Research and Development at Pacific Northwest National Laboratory (PNNL). PNNL is a multiprogram national laboratory operated for the U.S. Department of Energy (DOE) by Battelle Memorial Institute under contract no. DE-AC05-76RL01830. 

\section*{Data Availability}
Two datasets underlying this article, i.e. halo-based and correlation-based dark matter flow statistics, are available on Zenodo \citep{Xu:2022-Dark_matter-flow-dataset-part1,Xu:2022-Dark_matter-flow-dataset-part2}, along with the accompanying presentation slides 'A comparative study of dark matter flow \& hydrodynamic turbulence and its applications' \citep{Xu:2022-Dark_matter-flow-and-hydrodynamic-turbulence-presentation}. All data files are also available on GitHub \citep{Xu:Dark_matter_flow_dataset_2022_all_files}.

\bibliographystyle{Papers}
\bibliography{Papers}
\label{lastpage}
\end{document}